\documentclass[twocolumn]{aastex63}
\usepackage{graphicx}
\usepackage{float}
\usepackage{color}
\usepackage{url}

\usepackage{enumitem}
\usepackage{booktabs}
\usepackage{multirow}
\usepackage{enumitem}

\newcommand{\zhigh}{z_{\rm high}}

\newcommand{\LCDM}{$\Lambda$CDM}

\shorttitle{Origin of kinematic planes of satellites}  
\shortauthors{G\'amez-Mar\'in et al.}

\begin{document}

\title{
The origin of kinematically-persistent planes of satellite galaxies as driven by the early evolution of the local Cosmic Web in  $\Lambda$CDM }

	\author{Mat\'ias G\'amez Mar\'in}
	\affiliation{Departamento de F\'isica Te\'orica, Universidad Aut\'onoma de Madrid, E-28049 Cantoblanco, Madrid, Spain}

        \author{Isabel Santos-Santos}
        \affiliation{Institute for Computational Cosmology, Department of Physics, Durham University, South Road, Durham, DH1 3LE, UK}

	\author{Rosa Dom\'inguez-Tenreiro}
	\affiliation{Departamento de F\'isica Te\'orica, Universidad Aut\'onoma de Madrid, E-28049 Cantoblanco, Madrid, Spain}
\affiliation{Centro de Investigaci\'on Avanzada en F\'isica Fundamental, Universidad Aut\'onoma de Madrid, E-28049 Cantoblanco, Madrid, Spain}

	\author{Susana E. Pedrosa}
	\affiliation{Instituto de Astronom\'ia y F\'isica del Espacio, CONICET-UBA, 1428, Buenos Aires, Argentina}

    \author{Patricia B. Tissera}
    \affiliation{Instituto de Astrof\'isica, Facultad de F\'isica, Pontificia Universidad Cat\'olica de Chile, Av. Vicuña Mackenna 4860, Santiago, Chile}
    \affiliation{Centro de Astro-Ingenier\'ia, Pontificia Universidad Cat\'olica de Chile, Av. Vicuña Mackenna 4860, 7820436 Macul, Santiago, Chile}
    \affiliation{N\'ucleo Milenio ERIS - ANID}

    \author{M. \'Angeles G\'omez-Flechoso}
    \affiliation{Departamento de F\'isica de la Tierra y Astrof\'isica, Universidad Complutense de Madrid, E-28040 Madrid, Spain}
    \affiliation{Instituto de F\'isica de Part\'iculas y del Cosmos (IPARCOS), Universidad Complutense de Madrid, E-28040 Madrid, Spain}

    \author{H\'ector Artal}
    \affiliation{Tecplot, Inc., Rutherford, NJ 07070, USA}

\begin{abstract}

Kinematically-persistent planes of satellites (KPPs) are fixed sets of satellites co-orbiting around their host galaxy, whose orbital poles are conserved and clustered across long cosmic time intervals.  They play the role of `skeletons’, ensuring the long-term durability of positional planes.
We explore the physical processes behind their formation  in terms of  the dynamics of the local Cosmic Web (CW), characterized via the so-called Lagrangian Volumes (LVs) built up around  two zoom-in, cosmological hydro-simulations  of  MW-mass disk galaxy + satellites systems, where  three KPPs have been identified.
By analyzing the LVs deformations in terms of the reduced Tensor of Inertia (TOI),  we find an outstanding alignment between the LV principal directions and KPP satellites' orbital poles.
The most compressive local mass flows (along the $\hat{e}_3$ eigenvector) are strong at early times, feeding the so-called $\hat{e}_3$-structure, while the  smallest TOI axis rapidly decreases.  The $\hat{e}_3$-structure collapse marks the end of this regime and is the timescale for  the establishment of satellite orbital pole clustering when the Universe is $\lesssim$ 4 Gyr old. 
KPP proto-satellites aligned with $\hat{e}_3$ are those whose orbital poles are either aligned from  early times, or have been successfully bent at $\hat{e}_3$-structure collapse. 
KPP satellites associated to $\hat{e}_1$ tend to have early trajectories already parallel to $\hat{e}_3$. 
We show that KPPs can arise as a result of the $\Lambda$CDM-predicted large-scale dynamics acting on particular sets of proto-satellites, the same dynamics that shape the  local CW environment.

\end{abstract}







	\section{Introduction}\label{sec:intro}

                The discovery that satellites orbiting around the Milky Way (MW) and Andromeda (M31) have  a highly anisotropic distribution has been long considered to be one of the most-challenging issues for $\Lambda$CDM 
\citep[see][for a review]{Bullock17-review,Pawlowski18}.

Most MW satellites define a ``vast polar structure'' of dwarf galaxies with respect to the Galactic disk \citep[VPOS, see][]{Kroupa15,Pawlowski12,Pawlowski13}, with around 40\%  
of their orbital poles aligned with the normal vector to the plane of satellites \citep{Fritz2018}
\citep[see][hereafter PaperI]{SantosSantos20I},
suggesting that the VPOS is a robust positional structure and could be rotationally-supported.

Satellites orbiting around M31 have  also been  discovered to be  anisotropically distributed around it \citep{Koch06,McConnachie06,Metz07}, with almost half the population (15 out of 27 satellites)  forming a thin plane in positions, referred to as the ``GPoA'' or ``Great Plane of Andromeda'', and almost edge-on from our perspective. Moreover, in Paper I it has been shown that  a second positional plane, roughly  perpendicular to the former, shows up in M31.
Finally, positionally-flattened satellite structures  have also been detected in other major galaxies beyond the Local Group \citep[e.g.][]{Chiboucas13,Ibata:2014,Ibata15,Tully15,Muller16,Muller17,Muller18,Muller21,Paudel21,MartinezDelgado21, Heesters21}.
%

These recent observations have opened interesting debates on the issue of positional planes of satellites. Indeed, different authors have studied $\Lambda$CDM simulations 
finding that, while their presence is quite unusual \citep{Libeskind05,Libeskind09,Lovell11,Wang13,Bahl14,Cautun15,ForeroRomero2018}, positionally-detected planes of satellites can be found, in some cases even showing kinematic coherence similar to the planes found in the Local Group \citep{Gillet15,Buck15,Ahmed17,Maji17b,Garaldi18,Shao19,Samuel2021,Pham22,Forsters22}.


In many cases, however, these positional planes have been found to be unstable or transient structures, i.e., the membership of at least a fraction of satellites is lost within short timescales \citep{Bahl14,Buck16,Maji17b,Lipnicky17,ZhaoXinghai2023}. In line with these results, \citet[][hereafter Paper II]{SantosSantos2020II} have found that important fluctuations occur in the properties of positional planes as a function of time, presumably caused by the loss of a fraction of satellites that leave the planar structure on short timescales, while other transient satellites join it.
 These authors have also found that, at each simulation timestep,  only a fraction of satellites are in 
coherent co-orbitation\footnote{As in Paper I, II and III, in this Paper the term  co-orbitation will mean kinematic coherence,
\textit{ no matter the sense of rotation}, within an aperture $\alpha_{\rm co-orbit} = 36.87^\circ$, see \citet{Fritz18}.}
  within the planes, indirectly suggesting the possibility of a kinematic skeleton in positional planes.

This possibility has been explored by \citet{Santos-Santos_2023}, hereafter Paper III, who made a kinematic analysis of the satellite samples analyzed in Paper II from halo virialization time, T$_{\rm vir}$,  onwards. Specifically, two hydrodynamical, zoom-in MW-mass systems were studied. By focusing on the satellite orbital pole conservation and clustering, they identified the so-called kinematically-persistent planes of satellites (KPPs). These are groups of satellites, whose identities are the same 
along extended time intervals, and whose orbital poles are conserved and clustered in the same direction along them.
 In Paper III it was numerically shown that  KPP satellites,  on the one hand,
and satellites  members of the  thinnest  positional planes (i.e., those with  the lowest minor-to-major axes ratios $c/a$)
 among those with a  fixed  satellite fraction, on the other hand, share the same three-dimensional space configuration.
This result  numerically proves  that the positional planes  include  non-kinematically coherent satellites as well, i.e., satellites that are lost to the positional plane configuration,
and temporarily replaced by other transient-member satellites, as mentioned.
In this way, KPPs play the role of a kind of skeleton, shaping long-lasting planar structures, whose satellite membership fluctuates along time, except for the kinematically-coherent ones. 

We see that the key point of KPP structures is the clustering of a fraction of the host satellites' orbital poles and its persistence along time.
But the question remains about what is causing this clustering. Also, when is this clustering established. And whether or not the Cosmic Web (hereafter the CW) development has some role at answering to the two previous questions.

Indeed, the processes behind the clustering of orbital poles --and hence behind the origin of KPPs-- remain unclear.
Different approaches have been proposed in order to explain such phenomenon, such as group capture of satellites onto the central galaxy \citep{Lynden95,Li08,DOnghia:2008},
satellites being formed from tidal-dwarf galaxies in ancient gas-rich mergers 
\citep[][]{Kroupa10,Hammer:2013,Kroupa15}, capture of satellites during host mergers \citep{Smith16,Angus16}, 
or the effect of aspherical halo tides at increasing the  orbital pole collimation of satellites orbiting inside them, or at populating KPPs \citep[see, e.g.,][]{Shao19,Wang20}.

Additionally, partly based on early  observations of the alignment between the VPOS and the Large Magellanic Cloud
\citep[see][]{Lynden76,Kunkel76}, some authors \citep[e.g.][]{Samuel2021,Garavito-Camargo21} have suggested that the LMC infall onto the MW might help to explain the existence of the VPOS \citep[see, however,][]{Pawlowski21}. 
In Paper III it was shown that the late infall of a  LMC-like group of satellites is not needed in order to have kinematically-coherent satellite planes.
However, this presence might help to enhance the fraction of satellites in coherent co-orbitation, and more particularly,
the ratio of those rotating in one sense over those rotating in the contrary, within KPPs.

Although the previously mentioned processes might have
been operative along cosmic evolution and they could have concurrently contributed to satellite planes formation,
other authors' approaches have stressed the role of the CW evolution.
%
%
It is known that mass elements that currently form galaxies were  organized  at high redshift as a CW, whose emergence and evolution are analytically   described via the Zeldovich’s Approximation \citep{Zeldovich:1970} and its extension to the Adhesion Model \citep[see e.g.][and references therein]{Gurbatov:1989,Kofman92,Gurbatov:2012},
as well as via  numerical  simulations \citep[][and references therein]{Cautun:2014}.
These works show that the  morphology of the CW comes from a hierarchical, multiscale, anisotropic collapse, where  large-scale flattened structures, frequently containing coplanar filaments  \citep[see e.g.][]{AragonCalvo2010}  are but one of its elements.
 It is worth noting that when the collapse of a CW filament or a wall is mentioned\footnote{It is worth noting that the so-called collapse to a wall or to a filament appearing in the Zeldovich theory plus Adhesion Model (see references in the text) correspond to the formation of a caustic, a mathematical singularity where no mention is made to the statistical behavior of the particles they involve.}\label{foot:ZA_AM}, generally not a simple caustic formation is meant, but a multi-scale, complex, non-relaxed structure, made on its turn of  different smaller-scale CW elements and so on. We empirically know that, at a given scale, these are morphologically transient structures vanishing in favor of halos, where mass piles up.

%

There is an increasing evidence that the CW morphological development   shapes  some halo and galaxy properties.
Using numerical simulations \citet{Libeskind14,Kang15} have shown that the major axes of DM halos are  well aligned with the slowest collapsing directions  of the LSS density field Hessian (i.e., their major axes or filaments), 
see also \citet{Porciani2002} for proto-halo alignments.
 Other studies reach  similar conclusions analyzing simulations using different techniques \citep{VeraCiro2011,Shao16,Cataldi2023}. See also  \cite{Wang20} for similar results obtained in  the SDSS DR12 data set  analysis. 
Another basic example is  the role played by the CW
 in the acquisition of angular momentum by gas as it travels towards the galaxy formation region \citep{pichon11,Codis15,Kraljic20}.
According to the so-called Tidal Torque Theory 
\citep[TTT, see, e.g.,][for a review]{Peebles:1969,Doroshkevich:1970,White:1984, Schafer:2009},  angular momentum  acquisition at very high
redshift  
is the result of the misalignment between the Inertia Tensor and the Shear   Tensor due to tidal forces.
The extended TTT, revised  to include the CW anisotropic configuration \citep{Codis15,Kraljic20}, predicts 
that massive galaxies have their spin perpendicular to filaments, while for  low-mass galaxies   it is aligned with the filament.
These results have been confirmed through simulations, \citep[e.g.][]{Dubois14,Welker14} and observational data \citep[e.g.][]{Welker20}. 
Similar results, involving halos, are presented by \cite{Kang15} and references therein;  \citet{Codis12,AragonCalvo14,Welker14}. Alignment of halos with the filaments of the CW have been analyzed by \cite{GaVeena18,GaVeena19,GaVeena21} in different large-volume simulations.
  
Mass flows within  CW  structures also have an impact on the distribution of matter around galaxies and halos, especially on satellite distributions.
For example, some studies focus on the anisotropic character of subhalo / satellite capture along filaments \citep{Benjouali11,Libeskind14,Wang14,Kang15,Tempel15,Dupuy22},
as well as its possible consequences on satellite systems' shapes \citep{Tempel15,Wang20} or orbital  coplanarity \citep{Benjouali11,Goerdt13,Buck15}.
Other authors analyze alignments between the principal directions  of the inertia ellipsoids of satellite systems, on the one hand,  and those of a few-R$_{200}$-scale shells around their respective  host galaxies \citep{Shao16} or the axes of slowest collapse in the matter distribution at larger scales \citep{Libeskind15,Libeskind19}, on the other hand.
An analysis of DMO simulations led \citet{Libeskind12} to find  satellite orbital pole alignments with the intermediate  axis of the  shear field tensor \footnote{The shear tensor is the spatial rate of variation  of the deformation tensor.}. 
\cite{Welker18} study alignments of satellite galaxies  with filaments in their neighborhood in the Horizon-AGN simulation.



For our purposes here,  the interesting alignments involve KPP orbital poles  with directions 
characterizing large-scale mass flows converging towards  their host; specifically scales  large enough to include high redshift proto-satellites before they are bound to the host galaxy
and the mass surrounding them   responsible for their torquing  as well.  These kind of alignments  have not been fully addressed through  hydrodynamic  simulations yet, and are the key to unveil the physical origin of KPPs. 
Indeed,  satellites  were initially formed in connection with  galaxies, hence following the same  mass flows responsible for the formation of the latter (i.e., flows where satellites emerge essentially  as     
the nodes of mass distributions at smaller scales). 
As mentioned above,  flattened and elongated  structures on ever  larger scales  appear at particular locations as the CW develops.
Proto-satellites emerging  within these  regions
are expected to follow the mass flows causing them, traveling long distances before they reach the halo, 
 and suffering (to different extents) the effects of forces and torques coming from the CW (proto-)elements as they develop.
These effects would give rise to different types of satellite orbital pole alignments with the directions of main compression of flows, and possibly to orbital pole clustering.
This pole clustering  would explain the formation of KPPs.


In this paper this idea is explored making use of cosmological simulations.
%
Specifically, in this paper  we try to advance  in the quest for an answer to the   following questions, 
involving kinematic structuration of satellites  and their timescales: why and when was the clustering of KPP satellites  orbital poles established,
why not all of the satellites are involved in this clustering, 
and which role the Local Cosmic Web development played therein.

Regarding the analysis of kinematic organization, our methodology here will be
  based on previous works focusing on the study of Local CW developments around forming galaxies \citep{Hidding:2014,Robles:2015}.
 We aim at characterizing the local skeleton emergency by studying the shape deformation around galaxy-to-be-objects, quantifying the timescales of deformation, and
the possible changes of  the orientation of their principal directions and their freezing-out timescales.  This information will be used along with the orbital angular momentum information of those satellites orbiting at present time around these galaxies. By using the information about the kinematically-coherent persistent planes formed around these systems (see Paper III) we try to clarify whether satellites' anisotropical distribution forming satellite planes and the principal directions along which mass flows at larger scales are connected.

                        The paper is organized as follows: The simulations analyzed and their corresponding satellite samples are introduced in Sections 2 and 3, respectively. Section 4 is devoted to KPP satellites properties. The analysis of the mass density evolution around galaxy-to-be objects is addressed in Section 5, by introducing the Lagrangian Volumes (LVs) in order to characterize the Local Cosmic Web. Section 6 reports on the specific properties of the LV  evolution around the two galaxy systems analyzed in this paper. In Section 7 we study alignments of satellites (either individual poles or planes of kinematically-coherent satellites) relative to the LV’s principal directions across time.
Results are discussed in Section \ref{sec:Discussion}.  Finally, in the last section we  summarize our work and we expose the conclusions reached.




\section{Simulations}\label{sims}

The simulated satellite samples analyzed in this paper are the same as studied in Papers II and III, where the conditions to be met by galactic systems  are discussed, the codes, the runs  and the satellite identification methods are  described and the relevant references are given. To guide the reader, a brief summary follows.  

We study  planes of satellites orbiting around isolated, simulated 
host galaxies selected so that the following requirements are met:
(a), the host galaxy at redshift $z \sim 0$  is endowed with an extended ($R\sim15$ kpc) thin stellar
and gaseous disk;
(b), the assembly history of this central galaxy is free of major-merger events after halo virialization;
(c), the system  hosts a numerous ($\sim30$) satellite population around the host;
(d), the simulation is run with a resolution high enough to ensure that the analysis of  angular momentum conservation is made with sufficient accuracy. Thus, we require satellite objects to include more than 50 baryonic particles.

We have pre-analyzed a set of different  zoom-in cosmological  hydro-simulations  finding among them two that reach the previous prerequisites, the so-called Aq-C$^\alpha$  resimulated halo and  PDEVA-5004.
The two simulations make use of very different initial conditions, codes and physics prescriptions.
This fact will allow us to reach conclusions that are independent of simulation modeling.

\begin{table}
\centering
\scriptsize
\caption{Some parameters of the  cosmological model (C block), simulation (S block), host galaxy (HG block), and Lagrangian Volumes (LV block), both for the Aq-C$^\alpha$ and the PDEVA-5004 systems.  Gas particles whose temperature is higher than Temp$_{\rm lim}$  [K] are taken as 
pressurized, hot gas particles, and have not been considered here to calculate LV deformations. See text for parameter definitions.
}
\begin{tabular}{| l  |  l | c  |  c |}
\hline
 \multicolumn{1}{| c}{Block} &   \multicolumn{1}{c}{Parameter}		&\multicolumn{1}{c}{Aq-C$^\alpha$}      & \multicolumn{1}{c |}{PDEVA-5004}    \\
\hline
C  & $\Omega_m$              &        0.25                        &  0.28       \\
   & $\Omega_b$              &        0.04                        &  0.04        \\
   & $\Omega_{\Lambda}$      &        0.75                        &  0.72        \\
   & H$_0$ [km/s/Mpc]        &        73.0                        &  70.0        \\

\hline
S  & n$_s$                   &       1.0                        &      1.0    \\
   & $\sigma_8$              &       0.9                         &       0.81   \\
   & $m_{\rm bar}$ [M$_{\odot}$]     & $ 4.1\times10^5$          &  $3.94\times10^5$        \\
   & $m_{\rm dm}$  [M$_{\odot}$]     & $ 2.2\times10^6$          &  $1.98\times10^6$        \\

\hline
   &  {}                              &       $z$ = 0                         &       $z$ = 0  \\

\hline
HG & $r_{\rm vir}(z=0)$ [kpc]     &	241.26          	&	181.43                  	\\
   & M$_{\rm vir}$ [M$_{\odot}$]    &	     1.82$\times$10$^{12}$ &	3.44$\times$10$^{11}$		\\
   & $M_{\rm star}$ [M$_{\odot}$]   &          $8.6\times10^{10}$    &  $3.05\times10^{10}$       \\
   & $M_{\rm gas}$  [M$_{\odot}$]   &          $7.4\times10^{10}$    &  $8.6\times10^{9}$       \\
   & Temp$_{\rm lim}$  [K]          &          $2.0\times10^{5}$     &  $2.0\times10^{4}$       \\
   & T$_{\rm vir}$  [Gyr]           &      7                         &  6                         \\
   & T$_{\rm ta, halo}$   [Gyr]     &      4                          &  3              \\
\hline
LV       & $z_{\rm high}$             &     8.45                       &   10.00       \\
& & &  \\
K=10   & $R_{LV}$ [kpc]                 & 255.26                        & 166.07   \\
       & M$_{LV}$ [M$_{\odot}$]         & 3.15$\times10^{12}$      & 1.04$\times10^{12}$          \\

K=15   & $R_{LV}$ [kpc]                 & 382.89                    & 249.10   \\                    
       & M$_{LV}$ [M$_{\odot}$]         & 6.59 $\times10^{12}$      & 3.17$\times10^{12}$         \\

K=20   & $R_{LV}$ [kpc]                 & 510.52                    & 332.14   \\    
       & M$_{LV}$ [M$_{\odot}$]         & 8.11$\times10^{12}$     & 6.31$\times10^{12}$         \\

\hline
\hline

\end{tabular}
\label{tab:TabData1}
\end{table}


It is worth remembering that a standard two-phase process characterizes the halo mass growth: first a fast phase where mass growth
is largely provided by frequent merger activity, and then a slow phase, 
where growth rates and dynamical/merging activity are low.
The halo virialization time, T$_{\rm vir}$, roughly marks the separation between both phases.
In the slow phase  the system formed by the halo and its bound satellites evolves independently
from cosmic expansion.


\subsection{Aq-C$^\alpha$}

The initial conditions of this simulation come from the   Aquarius Project  \citep{Springel08},
    a selection of DMO  Milky Way-sized halos, formed in a \LCDM\ simulation run in a  $100 h^{-1}\, \rm Mpc$ side cosmological box.
    A new re-simulation of the so-called ``Aquarius-C" halo (hereafter Aq-C$^{\alpha}$),
    including the hydrodynamic  and subgrid models  described in \citet{Pedrosa15}, 
  \citep[see also][]{Scannapieco05,Scannapieco06},
    has been analyzed in this work.
    The initial mass resolution of baryonic and dark matter particles, $m_{\rm bar}$
    and    $m_{\rm dm}$, respectively, and the parameters of the  cosmological model are given in Table \ref{tab:TabData1} (C and S blocks).

    The  halo turn-around and  virialization happen  at a  Universe age of    T$_{\rm ta,AqC} \simeq$ 4  Gyr and   T$_{\rm vir,AqC} \simeq$ 7.0  Gyr, respectively. 
    In this case a 25\% of the halo mass is accreted after collapse.
    Properties of this host  galaxy measured at  the final redshift of $z=0$ are given in Table \ref{tab:TabData1} (Host Galaxy or HG block).

    This system  presents a quiet history  from $z\approx1.5$ to $z=0$, where no major mergers occur. 
By an age of the Universe of T$_{\rm uni} \sim $ 11.5 Gyr ($z=0.15$)
the main galaxy  captures a massive dwarf
($M_{\rm bar}=5.02\times10^{9}\,\rm M_\odot$), carrying its own satellite system (six members with the satellite identification criteria used in this paper).
The capture has been analyzed in Paper III, where it was shown that it   has no perturbing  effects on the
dynamical behavior of the rest of  Aq-C$^{\alpha}$'s satellite population. As a low redshift event, this capture  is beyond the scope of this paper.


\subsection{PDEVA-5004}

The PDEVA-5004 system comes from a zoom-in re-simulation made with the PDEVA code \citet{MartinezSerrano08} of a halo identified in a \LCDM\ run.
Parameters characterizing the cosmological model, the run, the host galaxy at $z=0$ 
and some characteristic timescales  are given in Table \ref{tab:TabData1}.  
For more details see  \citet{domenech12},  Paper II and references therein.

 The host halo  turn-around and  virialization events occur  at Universe age  T$_{\rm ta,5004} \simeq$ 3  Gyr and T$_{\rm vir,5004} \simeq$ 6 Gyr, respectively, a bit earlier 
than in the Aq-C$^{\alpha}$ system.
Only a 20 \% of the virial mass is assembled  after T$_{\rm vir,5004}$, and no major mergers show up in the slow phase.




\section{Satellite samples}
\label{satsamples}




\begin{figure*}
\centering
\includegraphics[width=0.48\linewidth]{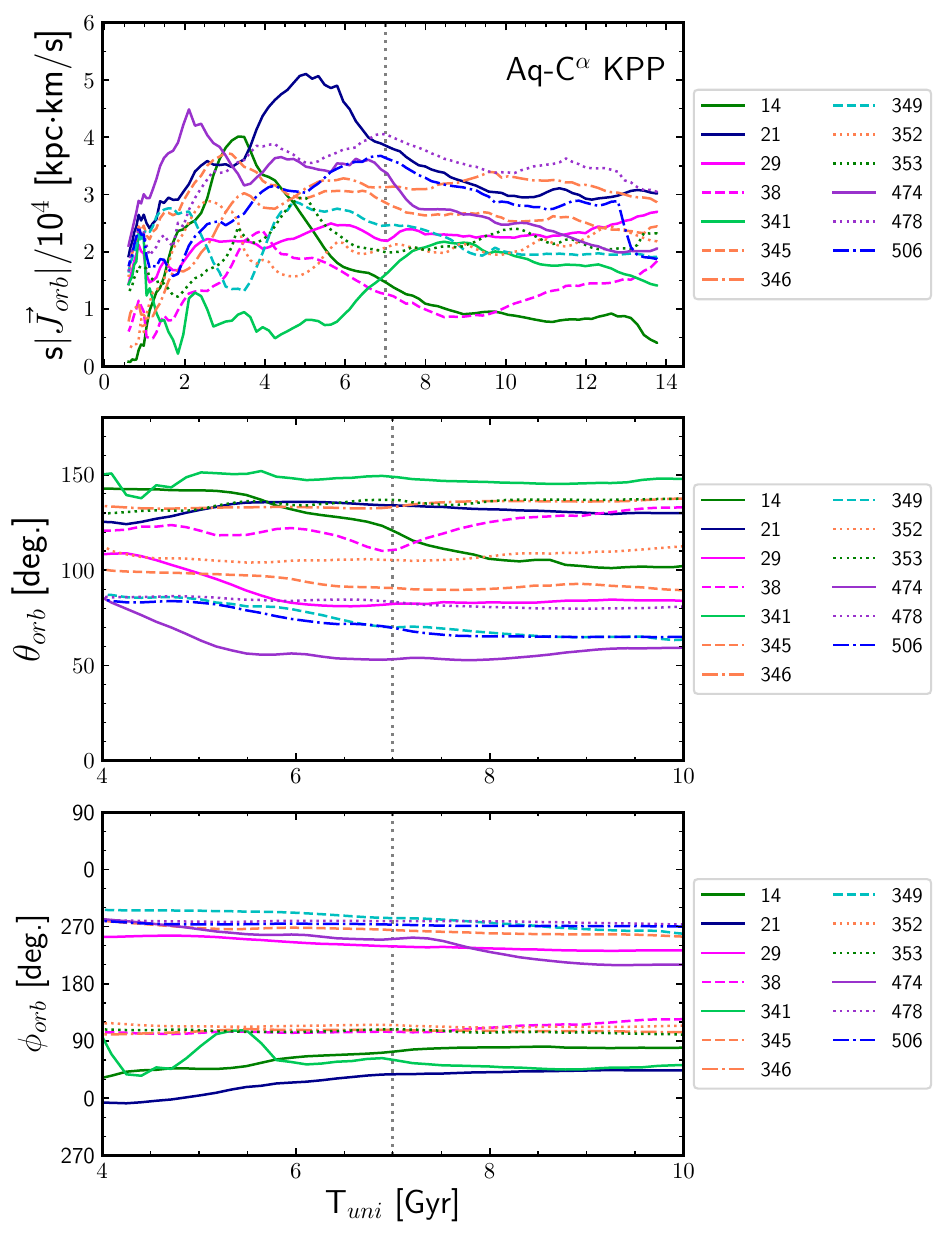}
\includegraphics[width=0.48\linewidth]{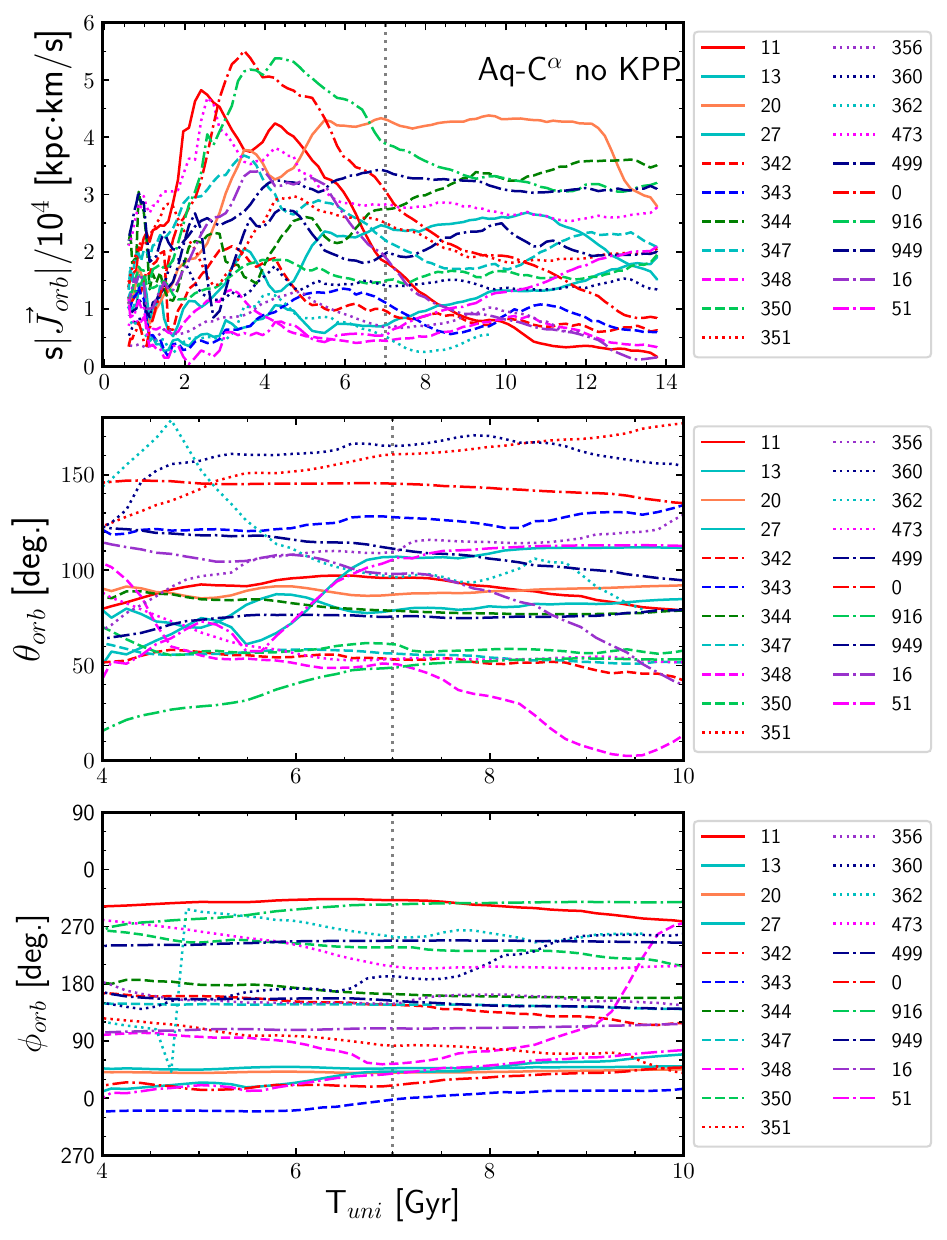}
\caption{ The components of the orbital angular momentum vector, $\vec{J}_{orb}$, of each satellite, as a function of time, for satellites in the Aq-C$^{\alpha}$ simulation, relative to axes that are kept fixed along cosmic evolution. The top panels show the evolution of the specific $\vec{J}_{\rm orb}$ vector moduli or magnitudes (hereafter $sJ_{\rm orb}$) from high redshift to $z$ = 0. The second and third rows show the $\vec{J}_{orb}$  directions, represented by the polar ($\theta$) and the azimuthal ($\phi$) angles in spherical coordinates. Their evolution is given
for T$_{\rm ta, AqC}$ = 4 Gyr $\leq$ T$_{\rm uni}$ $\leq$ 10 Gyr, an interval beginning at halo turn-around and centered at  the halo virialization time T$_{\rm uni} \sim$ 7 Gyr (marked with vertical lines).
The satellite samples are divided according to belonging (left panels) or not (right panels) to  kinematically-coherent, persistent planes, see Table 1, second column in Paper III.
Different colors and line types stand for the satellite IDs, as coded in the side-bar on the legends.}

\label{fig:sJ-zoom-pole-cons}
\end{figure*}



\subsection{Satellite identification}
   
The identification of satellites has been done, in both simulations, at two different times, i.e.
$z=0$ and $z=0.5$ ($T_{\rm uni}\sim 8.6$ Gyr),
in order to include satellites that may end up accreted\footnote{Satellite ``accretion'' stands for the 
disappearance of the satellite as an individual object, due to the partial  or total incorporation of the baryonic mass component into the central disk galaxy.} by the central disk galaxy and do not survive until $z=0$. 
We define satellite galaxies as those objects with stars ($M_{*}>0$) that are bound to the host galaxy within any radial distance. To ensure objects are bound we have computed their orbits back in time. 
The friends-of-friends algorithm and the SubFind halo finder \citep{SpringelWhite01} have been used to set structures and substructures in the Aq-C$^\alpha$ system,
\citep[see also][]{Dolag09}, while PDEVA-5004 satellites were selected using \texttt{IRHYS}\footnote{Simulation visualization and analysis tool developed by H.\,Artal.}. By tracing the particle IDs across snapshots, we have followed individual satellites back in time to times even before they were assembled as bound structures.
The total number of satellites is of 34 (35) in Aq-C$^\alpha$ (PDEVA-5004). Of these, 32 (26) survive until $z = 0$.
Satellites will be addressed
throughout the paper with an identification code (see
for example in Figures \ref{fig:sJ-zoom-pole-cons}  and \ref{fig:jorb_3Jorb}).

Satellites in Aq-C$^\alpha$ and PDEVA-5004 span baryonic mass ranges of $M_{\rm bar} = 8.5 \times 10^6 - 9.9\times10^8$ M$_\odot$ and $M_{\rm bar} = 3.9\times 10^7 - 1.8\times 10^8$ M$_\odot$, respectively (we recall that we impose satellites are resolved with a minimum of 50 baryonic particles).
The satellite mass distributions are addressed  in Papers  II and III, where it is shown that, 
at least in these two simulations and within the satellite mass range available here, the baryonic mass is not
a satellite property that determines whether a satellite belongs or not to the corresponding positional or  persistent plane.
This is an important result given our limited mass range  due to the current computational possibilities.


\subsection{Orbital properties}
\label{SatOrbProp}

Satellites present a diversity of orbital histories in both  simulations.  
While a small fraction of satellites end up accreted by the central galaxy's disk
(2 in the Aq-C$^\alpha$ system; 9   in PDEVA-5004),  most of them  have regular orbits with stable apocentric and pericentric distances. Some of them are backsplash satellites. Finally,  few satellite cases have just been captured by the main  galaxy's halo and have not yet had time to  complete  their first pericentric passage. 
Such late satellite incorporations have only been found 
in the  Aq-C$^{\alpha}$ system, where  several satellites 
show first  pericenters later than T$_{\rm uni}$  = 10 Gyr.
Conversely,  all of PDEVA-5004's satellites
are fully incorporated to the system by that time.


Figure \ref{fig:sJ-zoom-pole-cons} shows the evolution of the orbital angular momentum of satellites, $\vec{J}_{orb}$. Specifically, the top panels show
the  specific $\vec{J}_{\rm orb}$ vector moduli or magnitudes (i.e., normalized by their baryonic mass), hereafter $sJ_{\rm orb}$, for the Aq-C$^\alpha$ system, see Figure caption.
We can see that, for T$_{\rm uni}$ lower than $\sim$ 4 Gyr, most satellites' $sJ_{\rm orb}$ increases.
As mentioned in Section \ref{sec:intro}, the current paradigm sets that galactic systems acquire their angular momentum $\vec{J}$ at high redshift, prior to the system turnaround. 
After this moment, the lever arm becomes too low for an effective angular momentum gain, and $sJ_{\rm orb}$  stops increasing or even decreases. Most  satellites in Figure~\ref{fig:sJ-zoom-pole-cons}  show this behavior at high redshift. The same is true for the PDEVA-5004 satellite system.
Thus, the results of our simulations fit into the TTT scheme,  as far as the qualitative high-$z$ performance of some satellite's $sJ_{\rm orb}$  is concerned.
In the slow phase of mass assembly,  $s{J}_{\rm orb}$ is  roughly  conserved in many cases, while in others it is not.


The behavior of orbital poles after T$_{\rm vir}$ was studied in Paper III (see Aitoff projections in their figures 2 and B1). It was shown that the orbital  angular momentum directions  of most KPP satellites present  only modest changes with time after T$_{\rm vir}$, while for non-KPP members changes can be more relevant. 

For the purposes of this paper it is very relevant to analyze angular momentum conservation before T$_{\rm vir}$.
To have a deeper understanding of when pole conservation sets in, in the middle and bottom panels of Figure \ref{fig:sJ-zoom-pole-cons} we plot the orbital pole (polar and azimuthal angles with respect to axes that are kept fixed in time) evolution of KPP and non-KPP satellites within  the
T$_{\rm ta, AqC }\le$ T$_{\rm uni} \le 10$ Gyr interval, i.e., an interval beginning at halo turn-around time and centered at T$_{\rm vir}$.
Interestingly, some KPP-member satellites roughly maintain constant their orbital pole directions before T$_{\rm vir}$, while others change them only smoothly. 
Orbital pole changes are more frequent in non-KPP member satellites. We can also see that orbital pole clustering sets in already in the fast phase of mass assembly, an issue to be discussed in more detail in the next sections.


     


\section{Persistent Planes of satellites}
\label{PersistentPlanes}

\begin{figure}
\includegraphics[width=\linewidth]{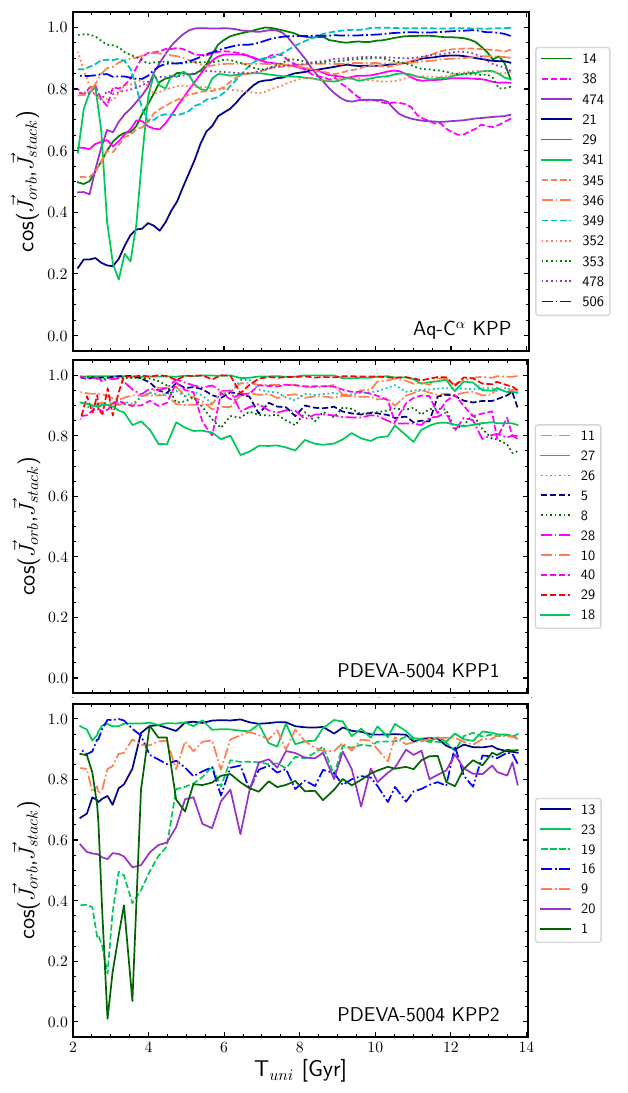}
\caption{Alignment between the axes of maximum  satellite co-orbitation, $\vec{J}_{\rm stack}$, and the orbital poles of satellites belonging to the KPP persistent plane in Aq-C$^{\alpha}$ (upper panel) and to  the two persistent planes KPP1 and KPP2 in PDEVA-5004 (middle and lower panels, respectively).
}
\label{fig:jorb_3Jorb}
\end{figure}




\subsection{Orbital pole clustering timescales}
\label{sec:KineSat}

As explained in Section \ref{sec:intro}, a   kinematically-coherent \textit{persistent} plane (KPP)  is  a fixed subset of satellites whose orbital poles are conserved  and remain  clustered for a long period of cosmic time, defining a high-quality planar structure in positional space in terms of a Tensor of Inertia (TOI) analysis \citep{Cramer}.
					 
In Paper III, axes of maximum satellite co-orbitation, $\vec{J}_{\rm stack}$,  were  determined  through the so-called ``Scanning of Stacked  Orbital Poles Method'' \footnote{Essentially, $\vec{J}_{\rm stack}$ defines a direction, fixed in time, around which a maximum number of satellite orbital poles in the  system cluster. It was defined in \cite{Santos-Santos_2023} using orbital poles for T$_{\rm uni}>$ T$_{\rm vir}$.}.
One (two) such axes have been found in the Aq-C$^\alpha$ (PDEVA-5004) systems, respectively,
giving 13  satellites for the Aq-C$^\alpha$ system KPP,   and 10 and 7 satellites, respectively,   for the two   PDEVA-5004 KPPs.

As an illustration of  these results, in Figure \ref{fig:jorb_3Jorb} we show  the time evolution of  $\alpha(\vec{J}_{\rm orb,i},\vec{J}_{\rm stack})$, the  angle formed by the axes of maximum co-orbitation and the orbital poles of each satellite member of the KPP planes. 
This Figure helps us answer to the following  important question: when is orbital pole clustering established? 
It is worth mentioning  that clustering  can appear at times much earlier than T$_{\rm vir}$,  before all the satellite members of a KPP are bound to the main galaxy.

According to Fritz's
co-orbitation criterion mentioned in Section \ref{sec:intro}, translated into
clustering properties, the $i$-th satellite orbital pole will
be considered to be aligned to the $\vec{V}$  vector when the
$\alpha(\vec{J}_{\rm orb,i},\vec{V})$  angle is smaller than  
$\alpha_{\rm co-orbit}=36.87^\circ$.
Figure \ref{fig:jorb_3Jorb} indicates that a fraction of satellites are aligned to their corresponding $\vec{J}_{\rm stack}$  already at T$_{\rm uni} =$ 2 Gyr. For those that are not aligned that early,  important changes in their poles direction occur in the  $\sim$ 2 - 4 Gyr time interval  for both simulations.
Specifically, an alignment timescale, T$_{\rm cluster, i}$ (relative to a given  $\vec{V}$ vector), can be defined
for the $i$-th satellite as the time when the $\alpha(\vec{J}_{\rm orb,i},\vec{V})$ angle reaches  a
cosine  higher than 0.8. Correspondingly, a measure of
the time when a given KPP sets in as a clustered bundle
of orbital poles around $\vec{V}$, T$_{\rm cluster, V}$, is given by the median of
the T$_{\rm cluster, i}$  values for each of its satellite members.
An  application of this protocol to data shown in    Figure \ref{fig:jorb_3Jorb}  gives
results for the median timescales (and
their corresponding differences with the 25-th  and 75-th
percentiles) of pole alignments with the $\vec{J}_{\rm stack}$ axes reported in Table \ref{tab:TimescaleTable}, T$_{\rm cluster, Jstack}$ entry.

We see that clustering sets in very early, especially in the case of the PDEVA-5004 system.
To understand why clustering sets in that early, we need to unveil its causes by analyzing LV deformations resulting from mass flows and their consequences. 



\begin{table}

\caption{Summary of timescales highlighted throughout the paper (values correspond to ages of the Universe, T$_{\rm uni}$/Gyr).  If a timescale is indicated in a specific figure, the figure number is indicated in parenthesis. Timescales in the first block apply to the Lagrangian volumes of each simulation. Times in the second block concern the satellite populations, and are given by medians and 25-75th percentiles when possible. For definitions, see text and timescales summary in Sec. \ref{sec:Timescales}.
} 
\scriptsize
%
\vspace{0.4cm}
\hspace{-2cm}
\begin{tabular}{|l | c  | c | c | c | c |}
\hline
\multicolumn{1}{|c|}{Timescale} & \multicolumn{2}{c|}{Aq-C$^{\alpha}$} &   \multicolumn{3}{c|}{PDEVA-5004}  \\
\hline
\multicolumn{1}{|c|}{T$_{\rm dir, e_3}$ (\ref{fig:direc-evol-DM})}   & \multicolumn{2}{c|}{$\sim$ 2.0} &   \multicolumn{3}{c|}{$\sim$ 0.5}  \\
\multicolumn{1}{|c|}{T$_{\rm shape, e_3}$ (\ref{fig:LVs-Properties_AqC-PDEVA})}   & \multicolumn{2}{c|}{$\sim$ 4.5} &   \multicolumn{3}{c|}{$\sim$ 3.5}  \\

\hline
\multicolumn{1}{|c|}{} &   \multicolumn{1}{c|}{KPP}& \multicolumn{1}{c|}{nKPP} & \multicolumn{1}{c|}{ KPP1 } & \multicolumn{1}{c|}{KPP2} & \multicolumn{1}{c|}{nKPP}    \\
\hline
T$_{\rm cluster, Jstack}$  (\ref{fig:jorb_3Jorb})     & 3.7$^{+ 0.9}_{- 1.5}$      &   --           & 2.2$^{+ 0.2}_{- 0.2}$ & 3.5$^{+ 1.0}_{- 1.1}$ & --  \\
T$_{\rm align, e_i}$      (\ref{fig:jorb_ei_AqC_PDEVA})     & 4.5$^{+2.5}_{-1.2}$               &   --           &  2$^{+0.2}_{-0.2}$           & 3.5$^{+0.5}_{-0.8}$ &    --  \\
T$_{\rm dist, plane}$     (\ref{fig:dist-median-perc-aqc})     & $\sim$ 3.5  &$\sim$ 3.5    &  $\sim$ 4.5           & $\sim$ 3.0   & $\sim$ 3.0  \\
T$_{\rm sat, infall}$             & 7.2$^{+ 1.8}_{- 3.9}$ & 8.8$^{+ 1.1}_{- 4.5}$ &         6.4$^{+ 1.4}_{- 1.2}$            &         4.5$^{+ 0.7}_{- 0.3}$              &   4.3$^{+ 0.5}_{- 0.7}$ \\
T$_{\rm sat, apo1}$               & 1.7$^{+ 1.3}_{- 0.2}$ & 3.8$^{+ 0.9}_{- 2.1}$ &           2.5$^{+ 0.5}_{- 0.3}$          &           2.4$^{+ 0.3}_{- 0.2}$            &  2.2$^{+ 0.2}_{- 0.2}$ \\
\hline
\hline
\end{tabular}
\label{tab:TimescaleTable}
\end{table}




\subsection{Evolution of the kinematic morphological  parameter }
\label{sec:krot}

The kinematic morphological parameter $\kappa_{\rm rot}$ is first  defined as the fraction of the kinetic energy of a given galaxy coming from ordered motions (i.e., in-plane and close to circular) relative to its  disk axis; see \citet{Sales:2012}.   Quantitatively this definition is:

\begin{equation}
\kappa_{\rm rot} = \sum_{i} w_{m, i} (v_{i,\phi} / v_i )^2
\label{krot}
\end{equation}
where $w_{m, i} = m_i/ M_s$ is the mass weight of the $i$-th constituent element  (with  $m_i$  its  mass  and  $M_s$  the mass of the system), and  $v_{i,\phi}$ and $v_i$  are the tangential
velocity relative to the system center of velocity in the plane normal
to the fixed  axis, and the modulus of the velocity of the
$i$-th constituent element, respectively.
$\kappa_{\rm rot}$  is a kinematic indicator of the morphology
        of a galaxy, such that those with  $\kappa_{\rm rot}$ lower than   0.5  are considered to be a
spheroid (dispersion-dominated galaxy), while galaxies
with higher values are rotation-dominated.

This definition can be easily extended to the set
of constituent elements of a given object or system, here
 satellite groups, relative to a fixed axis, here the  $\vec{J}_{\rm stack}$ axes.
Specifically,  we aim at studying  the evolution of the $\kappa_{\rm rot}$ parameter of KPPs as satellite sets versus that of sets of satellites outside these structures. In this kinematic analysis, the mass of each satellite is irrelevant and so we treat them as point sources, thus dropping the weighting term in equation \ref{krot}.

\begin{figure}
\centering
\includegraphics[width=\linewidth]{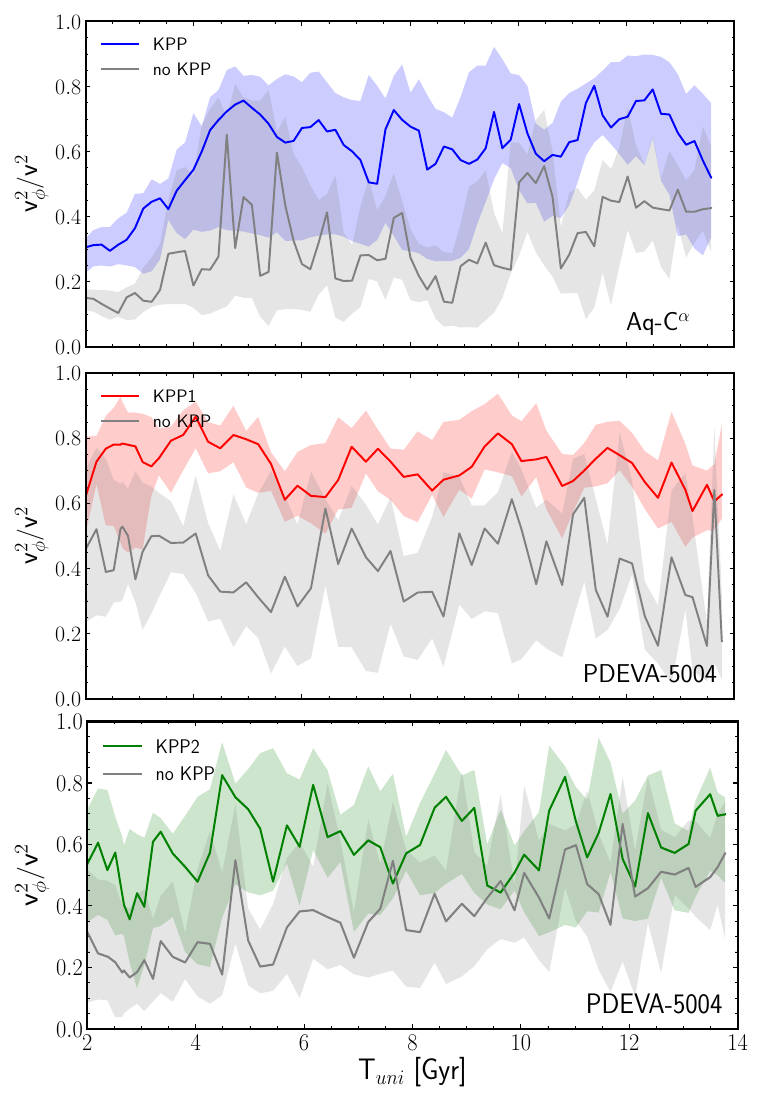}
\caption{ Behavior of the $(v_{i,\phi} / v_i )^2$ ratios as a function of cosmic time for  KPP and non-KPP satellites in the Aq-C$^{\alpha}$ and  PDEVA-5004 systems, see legends.
Lines are the median values at each simulation output time and the shaded bands give the corresponding 25-th and 75-th  percentiles.
Velocity components are taken relative to the corresponding $\vec{J}_{\rm stack}$ axes.}
\label{fig:Krot-median-satellites}
\end{figure}

Results for KPP planes are given in Figure \ref{fig:Krot-median-satellites}, where we show the median values (with their 25th-75th percentiles) for the $(v_{i,\phi} / v_i )^2(t)$ functions (note that the weighted mean is $\kappa_{\rm rot}$). A first remarkable result is the behavior of $\kappa_{\rm rot}$ as a function of time, with largely constant median values (fluctuations in time are  a result of low number statistics combined with parameter peaks at apocenter and pericenter). We see that satellites in KPP planes show high $\kappa_{\rm rot}$ values, and thus they behave as morphological disks  from a
kinematic perspective as well. 
Conversely, except for a few peaks, these medians are at any time lower  than 0.5 for satellites outside KPPs.
Finally, we see that high $\kappa_{\rm rot}$ parameter values set in early for satellites that will be KPP members, after  they increase at high redshift.
A  rough estimation of a timescale for the establishment of the maximum ordered disky motion, T$_{\rm max, rot}$, allowed by  Figure \ref{fig:Krot-median-satellites}, points towards 
T$_{\rm max, rot} \sim $ 5, and 4 Gyr, for the Aq-C$^{\alpha}$ KPP and   PDEVA-5004 KPP2 systems, respectively.
In the case of the PDEVA-5004 KPP1 system, satellites already show high $\kappa_{\rm rot}$ values since very early times, such that T$_{\rm max,rot}$ is virtually $<$ 2 Gyr. Indeed, as seen in Figure \ref{fig:jorb_3Jorb}, KPP1 satellites have their orbital poles aligned since very high redshifts, defining an in-plane motion.

Together with the results above on the timescales for clustering establishment, this  result here  reinforces the idea that satellite  kinematic coherence (that is, aligned orbital poles), and the appearence of disky-like orbits with  high $\kappa_{\rm rot}$  values (i.e., in-plane and circular motion), set in at very early times, when the proto-satellites are but part of the galaxy-to-be evolving environment, moving within it. Some answers to this possibility will be  given in the next sections.


\section{Setting up Lagrangian Volumes: Method}
\label{SettingLVup}
 
To study the local CW evolution,  we measure through TOI analysis the deformations of the Lagrangian Volumes (hereafter LVs) around the two host to-be-galaxies.
We define these regions by marking the particles that at high redshift are within a spherical volume around the center of the galaxy progenitor, and follow them forward in time (see \cite{Robles:2015}).  

\begin{figure*}
	\begin{center}$
	\begin{array}{cc}
	\hspace*{-0.6cm}\includegraphics[width=\textwidth]{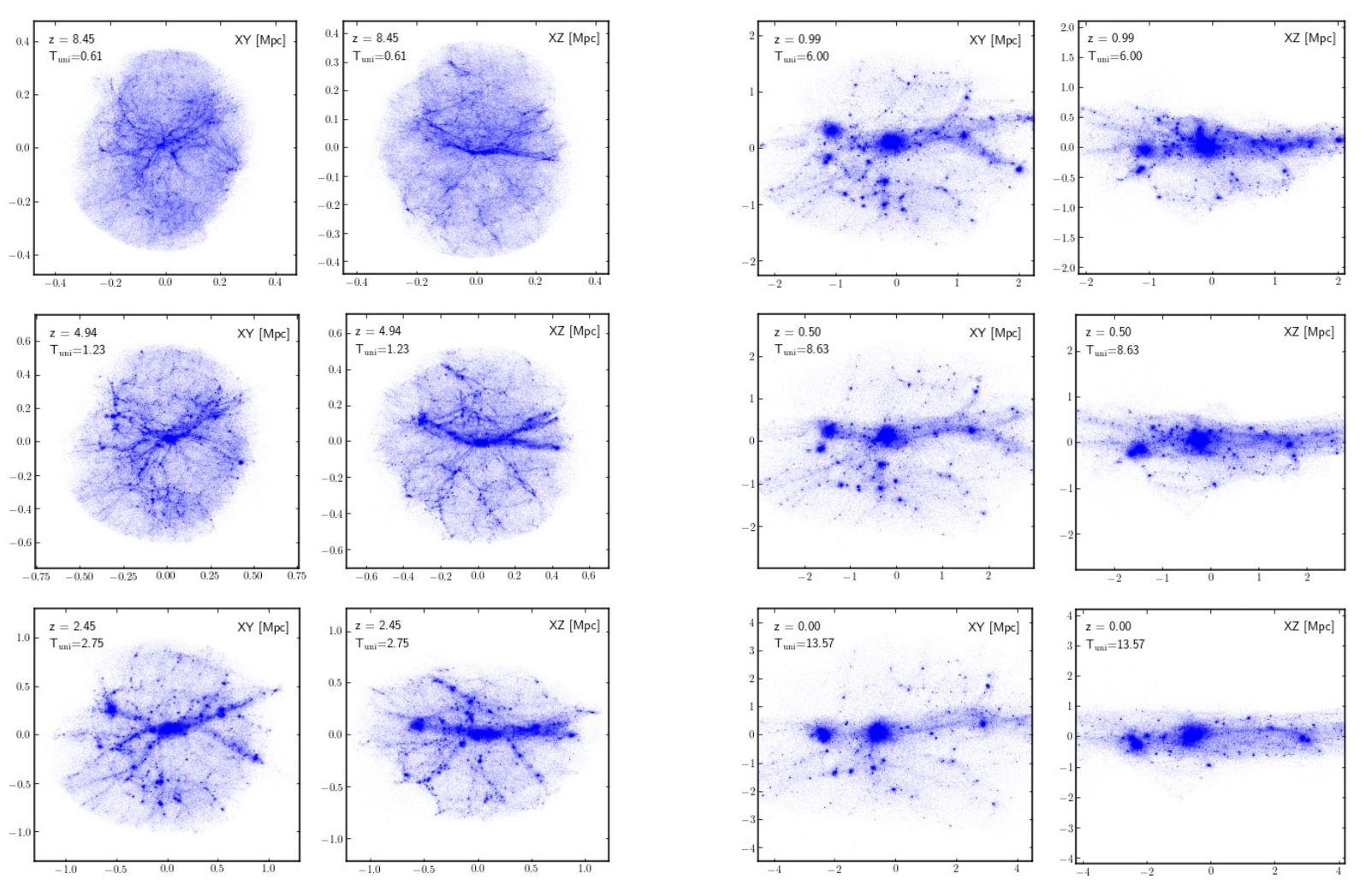}\\
	\end{array}$
	\end{center}
	\caption{ Shape evolution of the reference  Aq-C$^{\alpha}$ LV from $z_{\rm high}=8.45$ to $z$ = 0 (six snapshots). Two  LV projections along fixed axes  are shown for each snapshot, one along the $z=0$  $\hat{e}_3$ principal direction onto the XY plane (first and third columns), and the second one along the $z=0$  $\hat{e}_2$ principal direction onto the XZ plane (second and fourth columns). 
  The snapshot redshift $z$ and Universe age are given in each panel.   
This Figure shows how the initially spherical-like mass distribution flattens with time, in such a way that by $z \sim 1$ a wall-like structure has emerged, and that by $z \sim 0.5$ mass piles up in a predominant filament within the XY plane.
}
\label{fig:lagvol}
\end{figure*}

To build up  LVs, the first step is halo selection at $z = 0$.
Their  virial radii $r_{\rm vir,z=0}$ are given in Table \ref{tab:TabData1}. 
Next, for each halo at $z = 0$ we have traced back to high redshift  $z_{\rm high}$ (see values in Table \ref{tab:TabData1})   all the particles inside the sphere defined by its respective $r_{\rm vir, z=0}$.
Using the position of these particles at $z_{\rm high}$ we have calculated a new center of mass $\vec{r}_c$.
Then, we have selected at $z_{\rm high}$ all the particles enclosed by a sphere of radius 
$R_{LV} = K\times r_{\rm vir, z=0}/(1+z_{\rm high})$, with $K = 10, 15, 20$ (the motivation for this choice is discussed in  the next subsection),
around their respective centers $\vec{r}_c$
(see first row of Figure~\ref{fig:lagvol}).

These particles sample mass flows shaping the CW elements as the Universe evolves. Particles
follow geodesic trajectories until they possibly get stuck and begin the formation of,
or are accreted onto, a CW structure element (i.e., a caustic).
Our interest here focuses on the global  deformations of LVs, as an average of the fate of their constituent particles. Thus,
we have followed  these particles until  $z = 0$, i.e., we have followed the evolution  of the LVs from $\zhigh$ until $z = 0$. 
Note that, by construction, the mass of a LV is constant across evolution,  as well as the number of particles it is made of.

The choice of initially {\it  spherically} distributed sets of particles aims to unveil the anisotropic nature
of the local cosmological evolution, illustrated in  Figure \ref{fig:lagvol},
where the  LV corresponding to the Aq-C$^\alpha$  LV at $\zhigh$ and their corresponding deformations until its final shapes and orientation at $z = 0$ is  displayed.
In this Figure we note that the LV has evolved into a highly irregular, anisotropic and multiscale mass organization, including
very dense subregions as well as other much less dense and even rarefied ones,
with an overall flat structure from  $z \simeq 1$ onwards, corresponding to the formation of a large-scale sheet of the CW.
It is very remarkable that filaments become \textit{coplanar}, largely embedded into the flattening structure.

To quantify the local LV transformations
illustrated in Figure \ref{fig:lagvol}
we have calculated, at different redshifts,
   the reduced inertia tensor, $I_{ij}^{\rm r}$,  of each LV relative to its center of mass:
   \begin{equation}
   I_{ij}^{\rm r} =\sum_{n}m_n\frac{(\delta_{ij}r_{n}^2 - r_{i,n}r_{j,n})}{r_{n}^2}, \hspace{0.5cm} n=1, ..., N
   \label{reducedI}
   \end{equation}
where $r_{n}$ is the distance of the $n$-th LV particle to the LV center of mass and $N$ is the total number of such particles. 
We note that the summation does not include hot gas  particles, as their shapes are mostly driven by hydrodynamical / thermal pressure forces, see \citet{Robles:2015}.
We have used the reduced tensor instead of the non-reduced tensor \citep{Porciani2002a}
   to minimize the effect of
   substructure in the outer part of the Lagrangian volume \citep{Gerhard:1983,Bailin:2005}.
   In addition,
   the reduced inertia tensor is invariant under LV mass rearrangements in radial directions relative
   to the LV center of mass; this is, characterizations of the LV shape would not be affected by these mass flows, hence making
   the $I_{ij}^{\rm r}$ tensor particularly suited to describe anisotropic mass
   deformations as those predicted by the ZA and the AM and observed in Figure \ref{fig:lagvol}.

   In order to measure the LV shape evolution, first, we have calculated the principal axes of
	the inertia ellipsoid, $a$, $b$, and $c$, derived from
the eigenvalues ($\lambda_i$, with $\lambda_1 \leq \lambda_2 \leq \lambda_3$)
	of the $I_{ij}^{\rm r}$ tensor, so that $a\geq b\geq c$ , see \citet{GonzalezGarcia:2005,Robles:2015},

	\begin{eqnarray}
	a = \sqrt{\frac{\lambda_2 - \lambda_1 + \lambda_3}{2M}}, \qquad
	b = \sqrt{\frac{\lambda_3 - \lambda_2 + \lambda_1}{2M}}, \\ \nonumber
	c = \sqrt{\frac{\lambda_1 - \lambda_3 + \lambda_2}{2M}},
	\end{eqnarray}
	where $M$ is the total mass of a given LV. Note that $\lambda_1 + \lambda_2 + \lambda_3 = 2M$ and this implies $a^2+b^2+c^2=1$.
	We denote the directions of the principal axes of inertia by $\hat{e}_i$, $i=1,2,3$, where $\hat{e}_1$ correspond to the major axis,
	$\hat{e}_2$ to the intermediate axis and $\hat{e}_3$ to the minor axis.

The deformation of these Lagrangian Volumes is conveniently described by the triaxiality parameter, $T$, \citep{Franx:1991}, defined as: 

		\begin{equation}
		T = \frac{(1-b^2/a^2)} {(1-c^2/a^2)},
		  \end{equation}
		where $T=0$ corresponds to an oblate spheroid and $T=1$ to a prolate spheroid.
			An object with axis ratio $c/a>0.9$ has a nearly spheroidal shape, while one with $c/a < 0.9$ and $T<0.3$ has an oblate		triaxial shape. On the other hand, an object with $c/a < 0.9$ and $T>0.7$ has a prolate triaxial shape \citep{GonzalezGarcia:2009}.




\section{LV properties}
\label{LV_Prop}

As a simple characterization of the local CW dynamics around forming galaxy systems, we study the global evolution  of LVs from their initial spherical shape to the  structures they span at low redshift.
Different episodes stand out: i), orientation of principal directions and orientation  freezing-out timescales
(local CW skeleton emergency), 
ii), LV  deformations, and iii), characterization of the times when deformation slows down.
		They will be analyzed in turn.

As mentioned in Section  \ref{sec:intro}, the  local CW dynamics (or LV deformations) are mainly driven by the surrounding mass  density evolution. 
Thus, the scale the LV tracks must be big enough so that it not  only  traces back  the whole host-galaxy plus satellites system, but also to represent the local CW dynamics around them. 
We have tested the robustness of our results against changes in this scale by comparing our results using different LV radii 
$R_{LV} = K\times r_{\rm vir, z=0}/(1+z_{\rm high})$, with $K=10, 15$ (fiducial value) and 20 (see Appendix \ref{app:K_comparisons}). 
The results of the respective analyses have been compared.
Sizes and masses of the different LVs, taking into account their scale,  are given in Table \ref{tab:TabData1}.


\subsection{Evolution  of   the  principal directions}
\label{Pral-Direc-DMLV}

As said above, taking the LV as a whole, the $I_{ij}^{\rm r}$ eigenvectors, $\hat{e}_1(z)$, $\hat{e}_2(z)$ and $\hat{e}_3(z)$,  mark the directions of the major, intermediate and minor axes of its inertia ellipsoid at redshift $z$. 
It is very important to quantify the  changes in such directions as cosmic evolution proceeds.
 It is particularly important to find out whether or not the three eigendirections become fixed at a given time, say  T$_{\rm dir, e_i}$.  Should this happen, 
  mass rearrangements at LV  scales after T$_{\rm dir, e_i}$ would be  organized in terms of a ``skeleton'' or fixed preferred directions,
with the  $\hat{e}_3$ direction corresponding to that of maximum compression in the LV deformation, or the direction along which the overall mass flow has been maximum. 

\begin{figure}
\includegraphics[width=0.99\linewidth]{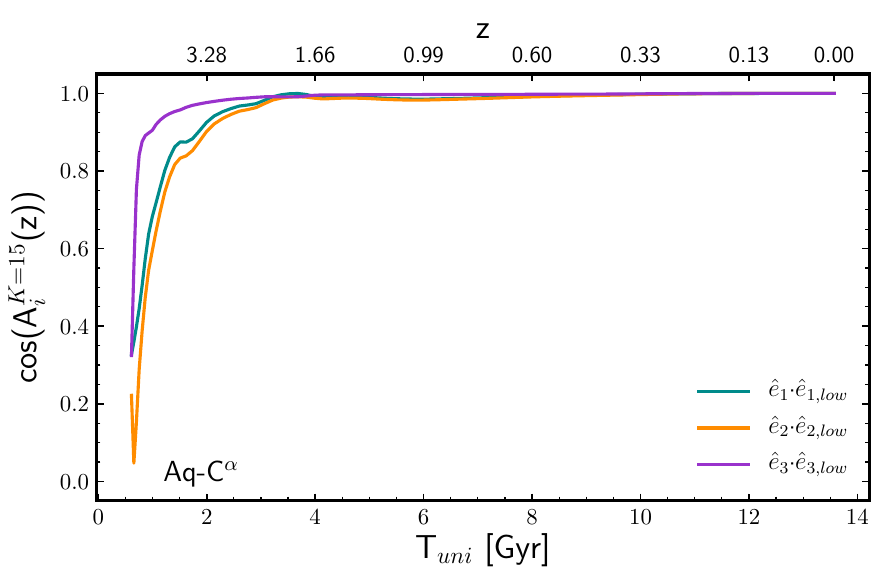}
\caption{ 
Evolution of the cosine of the angle $A_i$ formed by the eigenvectors $e_{i}(z)$ and $e_{i}(z=0)$ with $i$ = 1,2,3 for the  Aq-C$^{\alpha}$ simulation for the LV with a $K=15$ scale.
Upper horizontal axes give the redshift scale, while the lower ones stand for the Universe age T$_{\rm uni}$.}
\label{fig:direc-evol-DM}
\end{figure}

In Figure \ref{fig:direc-evol-DM}  we show the time evolution of $A_i(z)$, the angle formed by the eigenvectors
$ \hat{e}_i(z)$ and $\hat{e}_i(z = 0)$, with $i=1,2,3$ for the Aq-C$^{\alpha}$ simulation.
That is, we measure the deviations from the eigendirections at a given $z$ with respect to the final
eigenvectors for a LV scale of $K = 15$. Notice that only two out of the three $A_i$ angles are independent in such a way that if for instance $A_1=0$ then $A_2=A_3$.


We see that,  when $K=15$, $A_i(z)$ change  at high redshift. $A_3(z)$ smoothly  vanishes  by
T$_{\rm uni} \sim$ 2 Gyr $\equiv$ T$_{\rm dir, e_3}$. 
Results for PDEVA-5004 with $K=15$ are shown in Table \ref{tab:TimescaleTable}. Changes in the $\hat{e}_{3}$ and the other principal directions become unimportant very early,  except for a  small change of a 5\% at most in $\hat{e}_1$ and $\hat{e}_2$ occurring  around T$_{\rm uni} \sim$ 6 Gyr.
That is,  the LV deformations get their three  eigendirections fixed at high redshift, defining the freezing-out timescale T$_{\rm freeze}$ = 4 (2) Gyr for the Aq-C$^{\alpha}$  (PDEVA-5004) simulations
well before the systems enter the slow phase of mass assembly. 

It is important to figure out  whether or not this behavior depends on the LV scale.
In Appendix \ref{app:K_comparisons} (Figure \ref{appendix:direc-evol-DM}), we show that the evolution of the principal directions for a LV scale of $K=20$ for the Aq-C$^{\alpha}$ simulation makes it essentially  invariant under this change  in scale. The same results are obtained for the PDEVA-5004 simulation.
That being said, we proceed our analysis using a fiducial $K$ = 15 value.

\subsection{Eigenvalue and principal axes: Shape  evolution}
\label{sec:shapeevol}

\begin{figure}
\centering
\hspace{-0.4cm}\includegraphics[width=\linewidth]{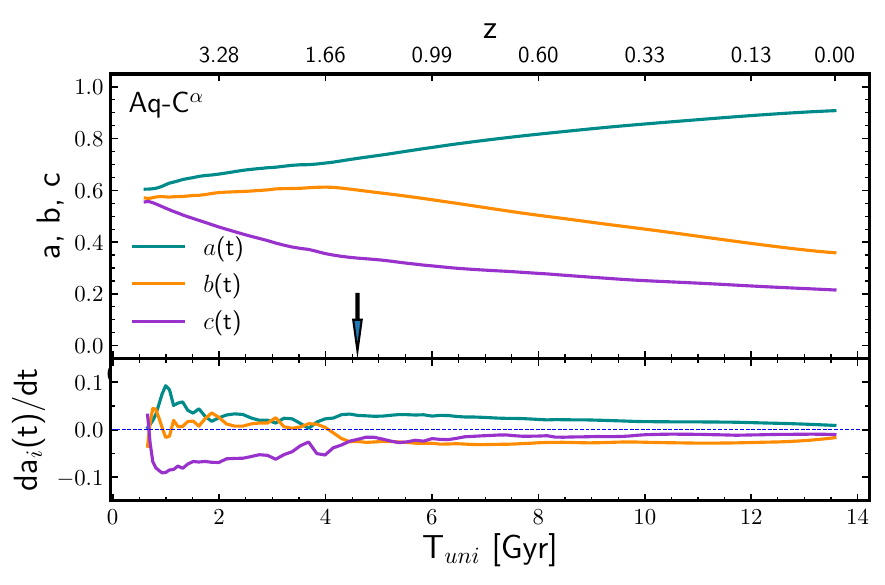}
\\
\hspace{-0.4cm}\includegraphics[width=\linewidth]{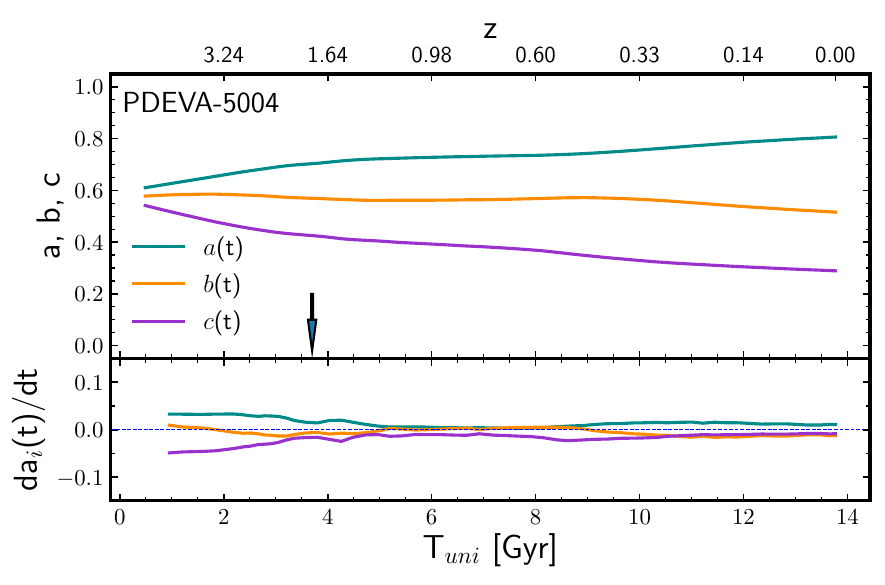}
\\
\hspace{+0.3cm}\includegraphics[width=\linewidth]{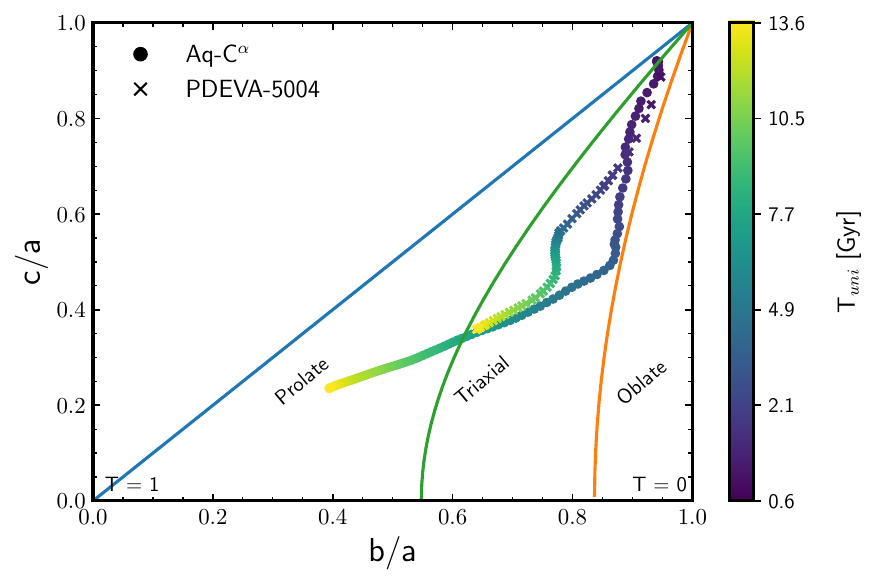}
\vspace{-0.5cm}
\caption{
Properties of the LVs of Aq-C$^{\alpha}$ (upper panel)  and PDEVA-5004 (middle panel) systems.			
Arrows mark the respective  T$_{\rm shape, e_3}$ timescales, when the rapid high redshift changes of the $c(t)$ principal axis slow down.  
Top and middle panels show the evolution of the principal axes lengths across time and their respective  derivatives.
The bottom panel gives the evolution of the axis ratios  $b/a$ and $c/a$ for both simulations, with the Universe age color-coded according to the colorbar. The $c/a$ versus $b/a$ plane is splitted into three regions, according to the values the $T$ shape parameter takes on them, see the blue, green and orange  iso-T curves.}
\label{fig:LVs-Properties_AqC-PDEVA}
\end{figure}

The shape evolution of the LVs  is presented in Figure \ref{fig:LVs-Properties_AqC-PDEVA}.
We show  the principal axes as a function of Universe age (results for Aq-C$^{\alpha}$ in the  top  panel, and PDEVA-5004 in the middle panel), and  their $b/a$ and $c/a$ ratios, color coded by the Universe age (bottom panel, see side-bar).

A remarkable result is the continuity of the $a(t), b(t)$ and $c(t)$ functions for all the LVs, with no mutual exchange of their
respective eigendirections across evolution, i.e., the local skeleton is continuously built up, in consistency with \citet{Hidding:2014} and \citet{Robles:2015}.
We see  that at high redshift the principal axes have  very similar lengths, as expected for a sphere.
Then, for both  LVs in these plots, the $a$  principal axes monotonously grows, while $c$  decreases
very rapidly and then the change slows down. Also, some  periods when the change is very slow occur in PDEVA-5004.
The $b(t)$  axis, corresponding to the $ \hat{e}_2$ principal direction, shows a decreasing behavior from T$_{\rm uni} \sim$ 4 Gyr onwards
in the Aq-C$^{\alpha}$ system, remaining almost constant until that moment. $b(t)$ shows some periods where it is almost constant in the  PDEVA-5004 simulations as well.

The overall shape deformation  of these LVs is  well quantified through the evolution of the $c/a$ and $b/a$ ratios and of  the triaxiality parameter, $T$.
The bottom panel of Figure \ref{fig:LVs-Properties_AqC-PDEVA} shows the $c/a$ versus $b/a$  diagram, where different shape specifications mark their corresponding parameter spaces, and the T=1.0, 0.7, 0.3 iso-T curves have been drawn.
%
We see that in both simulations, after a rapid evolution from a spherical shape ($b/a \simeq c/a \simeq 1$) towards more flattened structures 
($c/a$ decreases always, while $b/a$ is constant along some periods), shape changes slow down
as they increase their prolateness, with  the Aq-C$^{\alpha}$ system reaching a final shape more prolate than the  PDEVA-5004 system. Indeed, Figure \ref{fig:lagvol} shows explicitly how a planar structure in Aq-C$^{\alpha}$ is clearly formed by $z \sim 1.0$.
In the PDEVA-5004 simulation we note a fast change in $b(t)$ by  T$_{\rm uni}$ = 4 Gyr, marking a shape deformation from oblate to triaxial (see ``knee'' feature in the bottom  panel).

To better quantify how quickly the $a(t), b(t)$ and $c(t)$ functions change, their time derivatives are also  plot in the upper  and middle panel of Figure \ref{fig:LVs-Properties_AqC-PDEVA}.
We see fast changes in the $c(t)$ principal axis up to   T$_{\rm uni} \sim $ 4.5 Gyr  ($\sim 3.5 $ Gyr) for Aq-C$^{\alpha}$ (PDEVA-5004),
marked by arrows  in the top and middle panels as T$_{\rm shape, e_3}$ (see also Table \ref{tab:TimescaleTable}), and then 
the decrement rate becomes almost constant, and very low for PDEVA-5004 between T$_{\rm uni} \sim$ 5 and 8 Gyr. 
A rapid decrement in $c(t)$ reflects the strength of the early mass inflows in the $\hat{e}_3$ direction.
%
%
It is worth mentioning that any anisotropic mass inflow implies a mass rearrangement and, consequently, a change in the principal axes values. Figure \ref{fig:LVs-Properties_AqC-PDEVA} informs us when these anisotropic mass flows become unimportant.


An important point is the possible scale dependence of results on LV shape evolution. 
Our analyses indicate  that no remarkable, qualitative changes show up for either simulation  at the  larger scales tested here  in the evolution of their respective  principal axes lengths,  or the derivatives of the $a(t), b(t), c(t)$ functions.




\subsection{LV shape evolution from the perspective of CW structures formation and dissolution}
\label{sec:LV-CW-evolution}

An illustration of the LV global shape evolution for the Aq-C$^{\alpha}$ simulation just described is provided by Figure \ref{fig:lagvol}. 
This Figure shows that by redshift   $z \sim  1$ a flattened structure,  normal to the $\hat{e}_3$ principal direction
(hereafter, the $\hat{e}_3$-structure or, in this case, the $\hat{e}_3$-wall), clearly stands out for the first time in this plot. 
As explained in Section \ref{sec:shapeevol}, from this time onwards, vertical flows of LV material onto this flattened structure weaken to a great extent, while mass motions within it  still occur, as can be seen in the XY-plane  projections of the LV evolution. These mass motions lead, by $z \sim  0.5$, to the appearance of a prominent filament parallel to the $\hat{e}_1$ principal direction where most mass now piles up at the expenses of the wall-like structure population.

We note that the $\hat{e}_3$-structure is not a simple sheet, as the caustics appearing in the Zeldovich  model, but the result of a complex history of mass (smaller-scale CW elements) incorporations. 
Similarly, the filament in the $\hat{e}_3$-structure is not a simple filament. 
On its turn,  this predominant filament also  disappears later on in favor of halos (see $z=0$ panels of Figure \ref{fig:lagvol})  that form a prolate configuration, see Figure \ref{fig:LVs-Properties_AqC-PDEVA}, bottom panel.

The important point here is that the LV shape evolution expresses the CW unfolding at its  scale,  with all its complexity and diversity. 
The same is true in the case of the PDEVA-5004 system, with the difference that in this case the LV global shape is triaxial along almost all the evolution, leading to a prolate $\hat{e}_3$-structure (see lower panel in Figure \ref{fig:LVs-Properties_AqC-PDEVA}).




\section{LV alignments with satellite planes}
\label{sec:LV-alignments}

In this section we analyze alignments of satellites (either individual orbital poles or planes of kinematically-persistent satellites) relative to the LV's principal directions  across time, from $\zhigh$ to $z =  0$. 
The robustness of our results against $K$ changes is assured by  the previous discussion on scale effects in Section \ref{LV_Prop}.

\subsection{Alignments with the orbital poles of individual satellites}
\label{sec:alignOP}



\begin{figure*}												
\centering
\includegraphics[width=\linewidth]{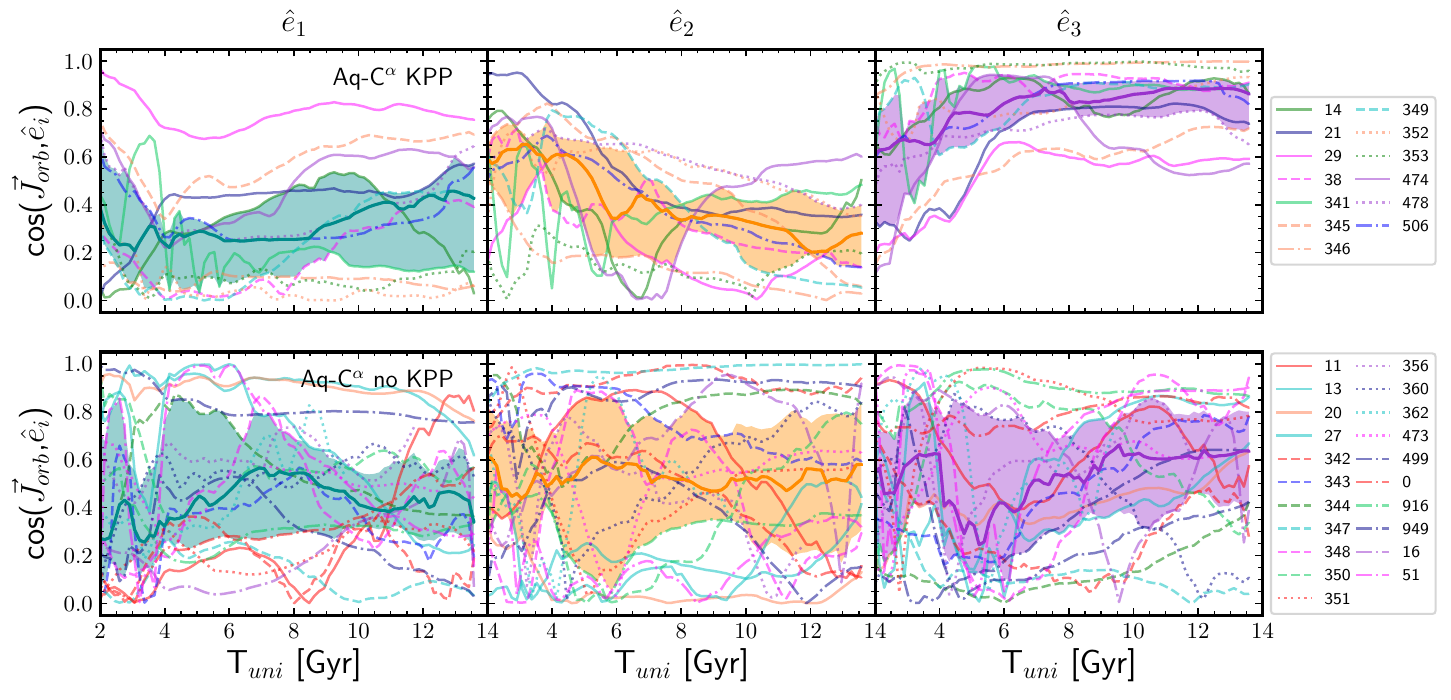}
\includegraphics[width=\linewidth]{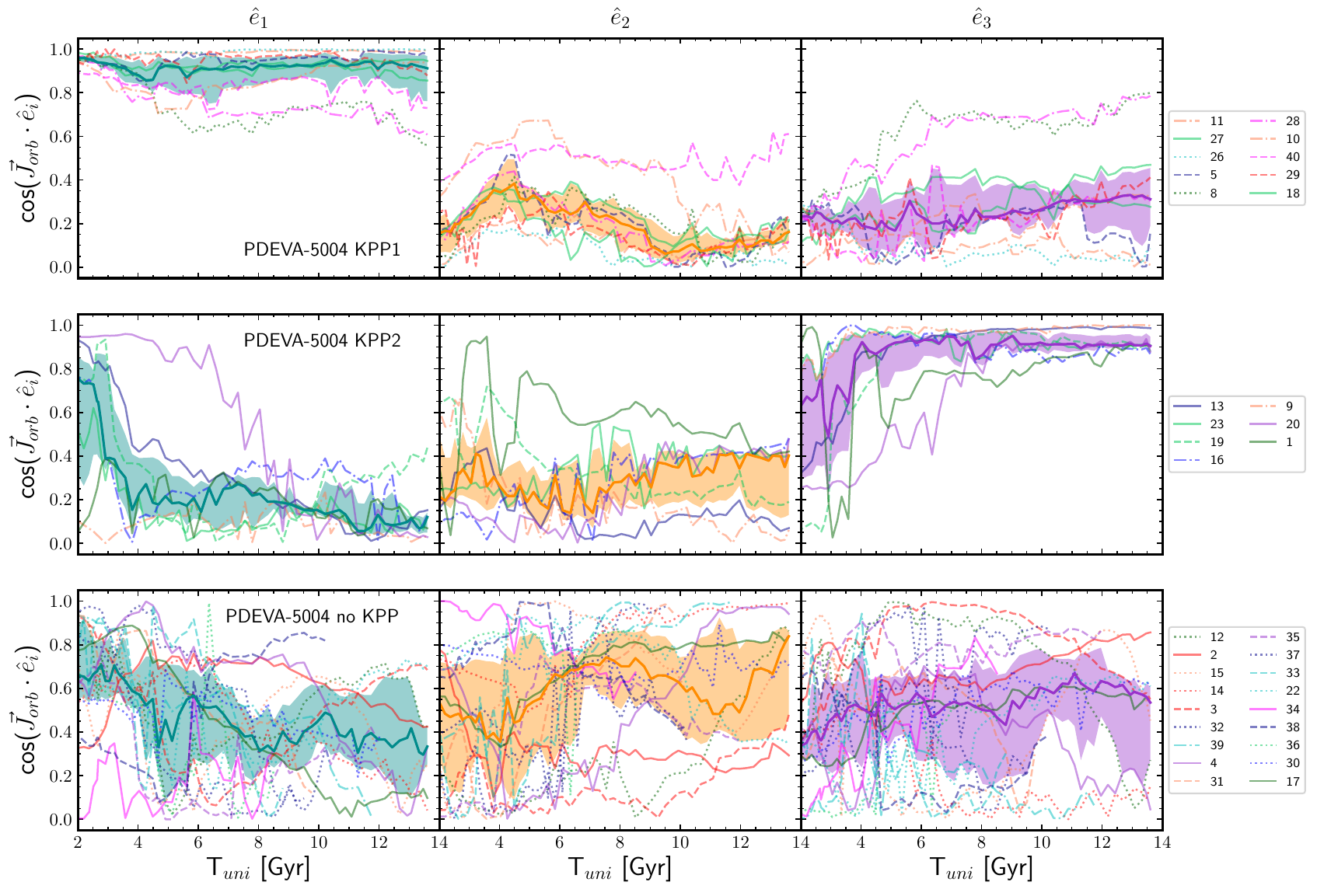}										
\caption{Alignment between the $\vec{J}_{orb}$ of satellites with respect to the principal directions of the LVs in Aq-C$^{\alpha}$ (upper block panels)
and PDEVA-5004 (lower block panels). 
Alignments are shown for satellites belonging to KPPs, and outside these structures as well, see legends.
Thin lines correspond to individual satellites, with colors and line-types as encoded on the right of the panels. 
Thick lines are the medians at each T$_{\rm uni}$, and shaded  bands mark the 25-75th  percentile range.} 
\label{fig:jorb_ei_AqC_PDEVA}
\end{figure*}


We first consider how the orbital poles of individual satellites are oriented relative to the  LV´s principal directions, 
$\hat{e}_1$, $\hat{e}_2$  and $\hat{e}_3$.


The time development of the alignments of orbital poles   with the principal axes  are given in Figures \ref{fig:jorb_ei_AqC_PDEVA}, both  for the 
Aq-C$^{\alpha}$ (upper block of panels) and PDEVA-5004 (panels in the bottom block) systems.
In this  Figure, panels in the first, second and third columns stand for the angle formed by the satellite orbital pole and the
principal directions $\hat{e}_1$, $\hat{e}_2$  and $\hat{e}_3$ of the LV reduced tensor of inertia. 
Thin lines correspond to individual satellite pole orientations, colored according to the satellite identity as given in the sidebars. 
  Cyan, orange and purple  thick lines represent the median values of the angle set at each timestep for $\hat{e}_1$, $\hat{e}_2$  and $\hat{e}_3$ alignments, respectively, while the shaded areas correspond to the respective 25-th  and 75-th percentiles. 

According to Figure \ref{fig:jorb_ei_AqC_PDEVA}, satellites in the  Aq-C$^{\alpha}$  KPP orbit on planes  close to normal to the direction of maximum global compression of the LV deformation, $\hat{e}_3$. That is, they move approximately  within  the flattened structure the initially spheroidal LV is deformed into along evolution.
Two satellites are already aligned by T$_{\rm uni} \sim$ 2. For those that are not,  changes in  their pole  directions 
leading to improved alignments mostly occur between T$_{\rm uni} \sim$ 2 - 4.5 Gyr.

A clear alignment signal (i.e., small angles) stands out for the PDEVA-5004 system, where the KPP1 satellite orbital poles tend to be close to parallel to the 
$\hat{e}_1$ axis. Indeed, the median of the $\cos(\vec{J}_{orb}, \hat{e}_1)$ is $\simeq$ 0.9 after T$_{\rm uni} \simeq $ 2 Gyr.  Therefore, after that,  KPP1 satellites tend to  orbit on planes close to normal  to the  $\hat{e}_1$ direction. 
%
On the other hand, satellite members of the KPP2 plane tend to have their poles  aligned with the $\hat{e}_3$ directions.
Thus,  KPP2 satellites  tend to orbit on planes close  to normal to the direction of maximum global compresion for the LV under consideration.
It is worth mentioning that the alignment improves (the shaded area becomes narrower) at lower redshifts for two out of three KPPs we have identified.

Most satellites outside any kinematically-coherent plane show no  alignment with either $\hat{e}_1$, $\hat{e}_2$ or  $\hat{e}_3$ directions.
They also show  a larger  spread in their angle values than those in the KPPs.

To extract information on the overall timescale for orbital pole alignment with the principal 
directions,  we look for the Universe age when the median values of the 
$\alpha(\vec{J}_{orb}, \hat{e}_i)$ angles (Figure \ref{fig:jorb_ei_AqC_PDEVA}), with $i$=3 for the Aq-C$^{\alpha}$ and the PDEVA-5004 KPP2 systems, and  $i$ = 1 for PDEVA-5004 KPP1 system, 
reach a value below $\alpha_{\rm co-orb} = 36.87^{\circ}$. 
Results are given in Table \ref{tab:TimescaleTable}, entry T$_{\rm align, e_i}$. We note the high dispersion of this value for Aq-C$^{\alpha}$ KPP. 
T$_{\rm align, e_i}$ gives also a measure of the timescale for the clustering of orbital poles. 
Results for T$_{\rm align, e_i}$ are consistent  with the corresponding  T$_{\rm cluster, Jstack}$,  
i.e., the timescale for the setting in of orbital pole clustering measured through alignments 
with J$_{\rm stack}$ (see Table \ref{tab:TimescaleTable}).


Another important point concerning satellite orbital pole alignments is how the probabilities of alignment with the axes of maximum co-orbitation, on the one hand, and with the principal directions, on the other hand, are related with each other.  
For the PDEVA-5004 simulation, the respective numbers of orbital poles aligned with the $\hat{e}_1$ and $\hat{e}_3$ principal directions are 8 (all of them aligned with the $\vec{J}_{\rm stack}$ axis defining the KPP1) and 9 (including the 7 satellite members of the  KPP2 group, aligned with the  $\vec{J}_{\rm stack}$ axis defining it).
Thus, a close relationship has been found in this case.
In the case of the Aq-C$^{\alpha}$ system, we have 13 satellites in the KPP aligned with the $\vec{J}_{\rm stack}$ axis, and 11  aligned with the $\hat{e}_3$ eigendirection.
The number of satellites aligned with both of them at the same time is 9. This gives for the conditioned fractions F($\vec{J}_{\rm stack}$ $\vert$ $\hat{e}_{3}$)         = 0.82, F($\hat{e}_{3}$ $\vert$ $\vec{J}_{\rm stack}$) =  0.69 and F( $\hat{e}_{i}$ $\vert$ no  $\vec{J}_{\rm stack}$) =   0.10.
Thus, we have found a  relatively high (low) probability that a KPP satellite has of having its pole aligned with the $\vec{J}_{\rm stack}$ axis, in case it is (it is not) aligned with the $\hat{e}_3$ or $\hat{e}_1$  principal directions. And conversely.

Summing up, our results in this subsection indicate that the physical processes leading to the local CW development, in the case of these two zoom-in simulations, might have a significant impact on the dynamics of satellites, shaping their trajectories before the satellites are gravitationally bound to the central galaxy. Indeed, the same processes that cause the evolution of the LV principal directions across time, induce the clustering of satellite orbital poles along some specific directions, hence contributing to the formation of kinematically-persistent structures.




\subsection{Global alignments with KPP planes}

\begin{figure}
\centering
\includegraphics[width=\linewidth]{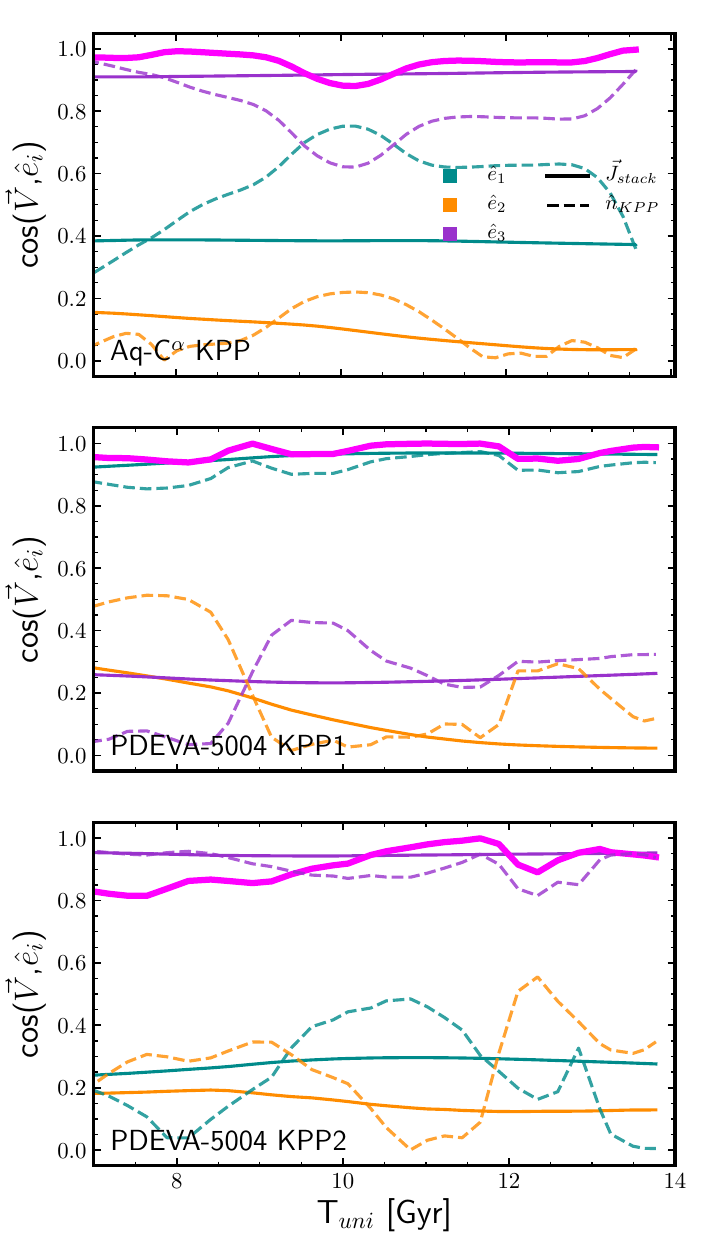}
\caption{Global alignments of the principal directions with $\vec{J}_{\rm stack}$ and the normal vectors to KPP planes, $\vec{n}_{\rm KPP}$, along cosmic evolution.
Upper panel shows the results for the Aq-C$^{\alpha}$ simulation, while middle and lower panels show the results for PDEVA-5004 KPP1 and KPP2, respectively.
Different line types stand for the cosine of the different angles.
Magenta thick lines: $\vec{J}_{\rm stack}$ and $\vec{n}_{\rm KPP}$.
Continuous thin lines: $\vec{J}_{\rm stack}$ and the principal directions.
Dashed  thin lines: $\vec{n}_{\rm KPP}$ and the principal directions.
Cyan, orange and purple colors  stand for each of the corresponding principal directions, as usual (see legends).
}
\label{fig:nKPPplanes_Jstack}
\end{figure}

To characterize KPP orientations relative to the principal directions, either their normals, $\vec{n}_{\rm KPP}$, or the respective axes of maximum co-orbitation, $\vec{J}_{\rm stack}$, can be used. These are not equivalent analyses, as the normals $\vec{n}_{\rm KPP}$ are returned by the ToI analyses of the KPP satellite positions, while $\vec{J}_{\rm stack}$ determination  uses the full 6-dimensional phase space information.
The angles between both vectors are given in Figure \ref{fig:nKPPplanes_Jstack} (magenta thick lines), where we see that these vector are highly aligned, except for the PDEVA-5004 KPP2 plane, more noisy. We also see that, as expected, $\vec{n}_{\rm KPP}$ and $\vec{J}_{\rm stack}$  are not perfectly parallel.



Information on the alignments among the principal directions of LVs and co-orbitation axes, $\vec{J}_{\rm stack}$, 
is given in Figure  \ref{fig:nKPPplanes_Jstack} (thin continuous lines) from T$_{\rm uni}$ = 7 Gyr onwards, both  for the Aq-C$^{\alpha}$ and the  PDEVA-5004 systems, see legends. 
As expected from the previous subsection, the $\vec{J}_{\rm stack}$ axis forms a small angle with the $\hat{e}_3$ direction in the case of Aq-C$^{\alpha}$ KPP, and PDEVA-5004 KPP2 along most of the period analyzed. In other words, the direction of maximum co-orbitation in these two cases is close, across time,  to that of maximum compression of the LV of reference here. 
The $\vec{J}_{\rm stack}$ axis for PDEVA KPP1 plane, on its turn, is close to the $\hat{e}_1$ direction.

Informations on KPPs  orientations relative to the LV principal directions can be obtained from their respective normal vectors
as positional planes, $\vec{n}_{\rm KPP}$.
Figure \ref{fig:nKPPplanes_Jstack} shows the angles between  $\vec{n}_{\rm KPP}$  and the LV principal directions (dashed thin lines, see legends).
Results for the PDEVA-5004 system show that the KPP1 (KPP2) structures, when considered as positional planes, are aligned with the $\hat{e}_1$ ($\hat{e}_3$) 
principal directions.
For the KPP identified in Aq-C$^{\alpha}$, its normal  vector aligns with $\hat{e}_3$ at high $z$ and close to $z = 0$. 
In between the alignment dims, and the angles both vectors form are similar, along some time intervals, to that $\vec{n}_{\rm KPP}$ forms with the $\hat{e}_1$
principal direction.  			

	Put together, results shown in Figure \ref{fig:nKPPplanes_Jstack}  indicate that the  local environment flattening  
correlates with satellite kinematics rather than with  their space positions.
Indeed, kinematically coherent  KPP satellites are   characterized by their respective  $\vec{J}_{\rm stack}$ axes, axes that highly align with the principal directions of the  LVs, while this alignment  gets worse when the positional planes  (as characterized by their respective normals, $\vec{n}_{\rm KPP}$) are considered.




\section{Discussion}
\label{sec:Discussion}


\begin{figure*}
\centering
\includegraphics[width=\linewidth]{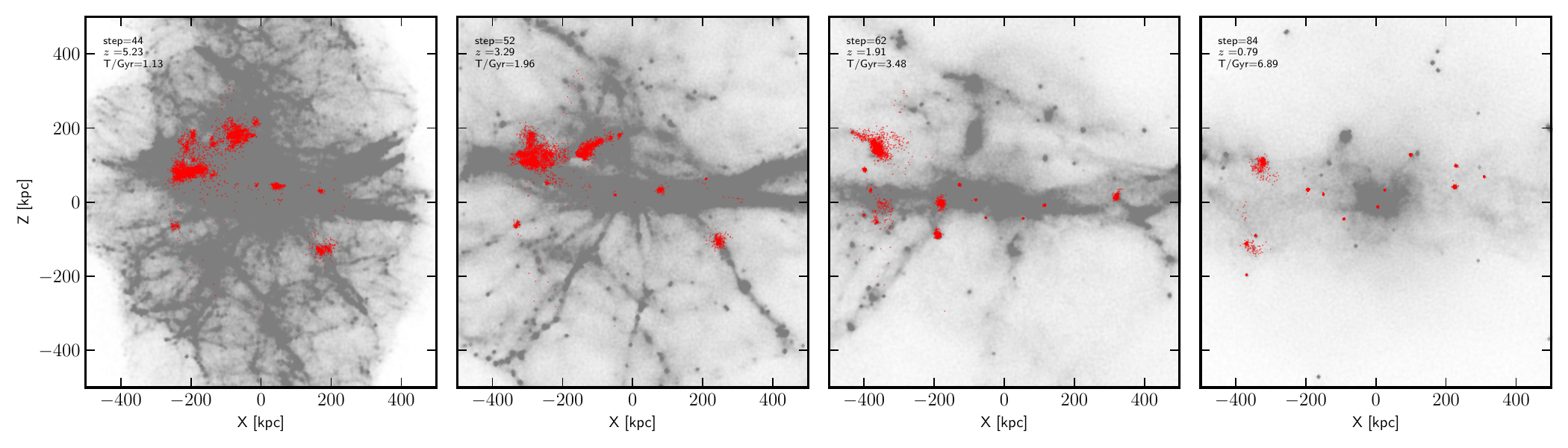}
\caption{Projections along the $\hat{e}_2$ principal direction of the density field around the host galaxy formation site for different values of the Universe age given on the top left corner of each panel (Aq-C$^{\alpha}$ simulation in physical coordinates). Red points sample the (proto-)satellite mass elements.
Evolution goes from left to right panels, and the rightmost panel refers to the host halo virialization time.
}
\label{fig:Proj-Dens-Sat}
\end{figure*}

\begin{figure}
\centering
\includegraphics[width=\linewidth]{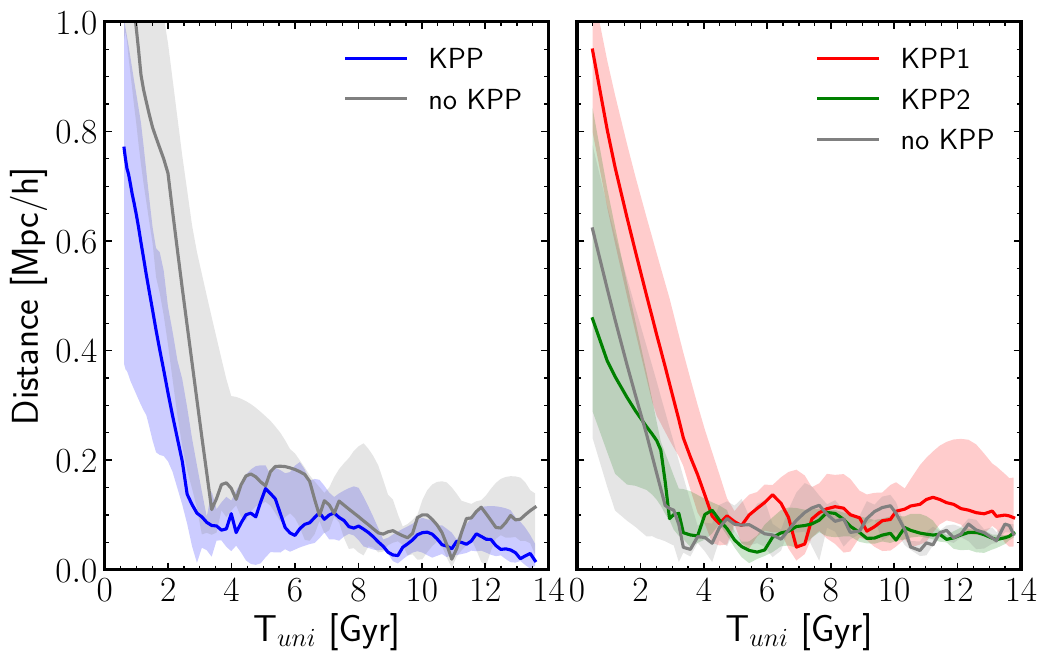}

\caption{Median distances of the different satellite populations in the Aq-C$^{\alpha}$ (left panel) and the PDEVA-5004 (right panel) simulations  to the corresponding plane normal to $\hat{e}_3$ that contains the center of mass of the LV.
Shaded areas stand for the corresponding 25-th and 75-th percentiles.}
\label{fig:dist-median-perc-aqc} 
\end{figure}

\subsection{A CW wall-like structure in Aq-C$^{\alpha}$}

   
As suggested by Figure \ref{fig:jorb_ei_AqC_PDEVA}, the $\hat{e}_3$-structure plays an important role as driver of orbital pole alignments in the two simulations analyzed here. 
Its formation and fate has been addressed in subsection  \ref{sec:LV-CW-evolution}, in terms of the mass flows feeding it and, later on, piling up mass in a predominant filament that finally  fades away in favor of halos.  
Let us now analyze the possible links between these processes and satellite plane formation. 
We  do so by following the overall mass flows feeding the $\hat{e}_3$-structure from either side, using the satellites selected in this work as markers of the flows. Note that, in the following, we will use the term ``$\hat{e}_3$-plane'' (in contrast to ``$\hat{e}_3$-structure'') when referring to the mathetmatical plane that is perpendicular to the $\hat{e}_3$ direction and contains the center of mass of the LV. 

Figure \ref{fig:Proj-Dens-Sat}  shows some snapshots of the  joint CW and KPP (proto-)satellite\footnote{In this paper, the term (proto-)satellite is used when both proto-satellites and satellites are meant.} evolution in physical coordinates, in a reference frame with the z-axis aligned with $\hat{e}_3$, the y-axis aligned with $\hat{e}_2$, and the x-axis aligned with $\hat{e}_1$, centered on the host galaxy formation site.
We see that satellite-to-be mass elements (red points in the Figure) come from small, dense sub-volumes, that at high redshift are small walls or filaments. 
Satellites are eventually formed by feeding from these mass elements. In some cases these mass elements disappear in favor of collapsed satellites, which in these cases are essentially born in isolation. For example the red structure located at $Z\sim 180$ kpc and $X\sim -200$ kpc (in the first panel) evolves into a unique, isolated satellite.
In other cases, satellites follow the filament they are formed from in a wide sense.
These snapshots illustrate  the idea that  the global effects of the forces and  torques produced by  the whole CW on a particular (proto-)satellite  drives its incorporation to the $\hat{e}_3$-wall  in the Aq-C$^{\alpha}$ simulation,  with the orbital pole alignment results shown in Figure \ref{fig:jorb_ei_AqC_PDEVA}.

%

The question arises of why some satellites are not part of KPPs. 
As already mentioned, in Figure \ref{fig:Proj-Dens-Sat}    we see that  different satellites reach the $\hat{e}_3$-wall under different circumstances: some flow through filaments along with mass elements that eventually feed into the formation of the $\hat{e}_3$-structure,  others are part of a small-scale ‘wall’ that merges with the $\hat{e}_3$-structure in a parallel manner, and finally others reach the $\hat{e}_3$-structure freely, having cannibalized  along the way the mass of the  CW element they were initially embedded in.  The natural evolution of the CW dynamics therefore leads to satellite trajectories with different characteristics as they reach the $\hat{e}_3$-structure (e.g. distance to the $\hat{e}_3$-plane, angle with respect to the plane), which in turn affects how the orbital pole alignment for each particular satellite comes about.


Qualitatively, we observe that satellites with trajectories that initially are closer to the $\hat{e}_3$-plane tend to run obliquely  relative to it  and are smoothly captured by the wall.  
Some satellites are almost in-plane from high redshift.
On the other hand, satellites coming from further away fall onto the $\hat{e}_3$-plane at higher velocities, more perpendicularly to it, and  they tend to maintain a velocity component normal to the $\hat{e}_3$-plane, such that their orbital poles are contained within this plane. In addition, satellites with a low impact parameter (i.e. low minimal distance with respect to  the proto-host galaxy center of mass) never suffer important specific angular momentum $sJ_{\rm orb}$  gains (see Figure \ref{fig:sJ-zoom-pole-cons}), resulting in small apocenter orbits,   with a high probability of suffering disturbing effects  from the central regions of the system, and so changing their orbital pole directions.

We have studied the evolution of distances from each satellite to the $\hat{e}_3$-plane.
The left panel in Figure \ref{fig:dist-median-perc-aqc} shows the time evolution of the median distances for the different satellite populations in Aq-C$^{\alpha}$, together with the corresponding 25th - 75th percentiles (in comoving coordinates, i.e., removing the effects of the background Universe expansion).
%
Two different regimes are clearly visible, either in the KPP, or non-KPP populations: 
At early times, distances decrease rapidly, an indication of  strong mass flows in the $\hat{e}_3$ principal direction, feeding the $\hat{e}_3$-structure.
In a second phase, median distances show  fluctuations, with essentially constant amplitude (in physical coordinates), except for  non-linear effects around $\sim$ 10 Gyr due to a pericenter accumulation event, affecting mainly the non-KPP population behavior. This constancy indicates that satellites do not feel the background expansion anymore.
The two regimes define a separation interval of time,  T$_{\rm dist, plane}$, whose specific values depend on the satellite population, see below. For the sake of clarity, we define this value as the time when the distance curve reaches a minimum for the first time.
%
This behavior, namely a rapid distance decrement  leading to a system with roughly constant size, decoupled from the background expansion, is reminiscent of  collapse events  suffered by halos as described by the Spherical Collapse Model \citep{Padmanabhan93} plus the ensuing violent relaxation \citep{LyndenBell1967}  process leading to an equilibrium configuration. In the absence of any theory or model to describe the statistical fate of the particles involved in the collapse towards a wall (see footnote \ref{foot:ZA_AM}), we focus on  the empirical characteristics just mentioned of the $\hat{e}_3$-structure behavior in a time interval around $\sim$ 4 Gyr, and  due to the reminiscences found with some characteristics of halo collapse in 3D, we will hereafter refer to this event as `$\hat{e}_3$-structure collapse'.

\begin{figure*}
\centering
\includegraphics[width=0.42\linewidth]{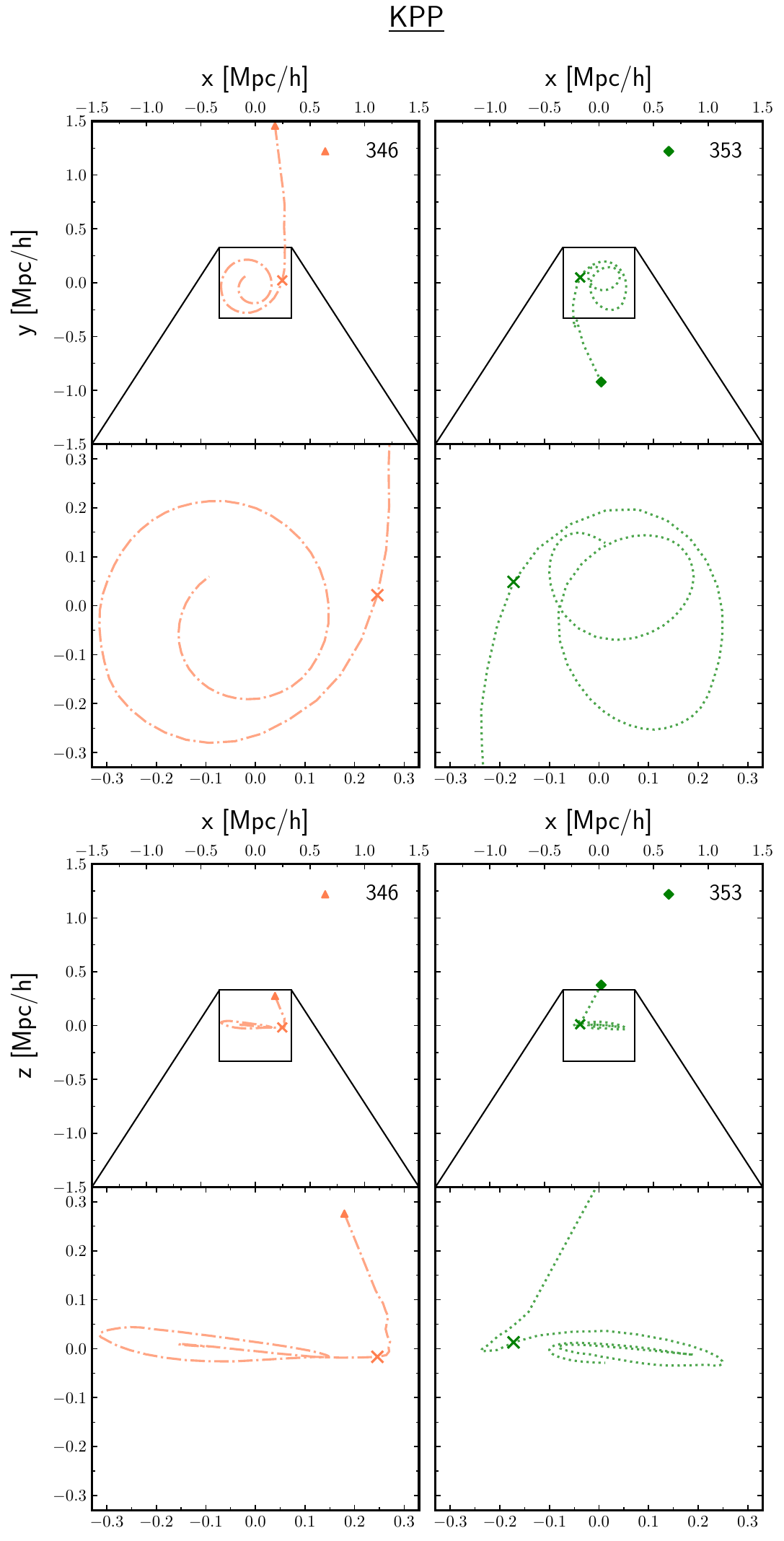}
\hspace{0.5cm}
\includegraphics[width=0.42\linewidth]{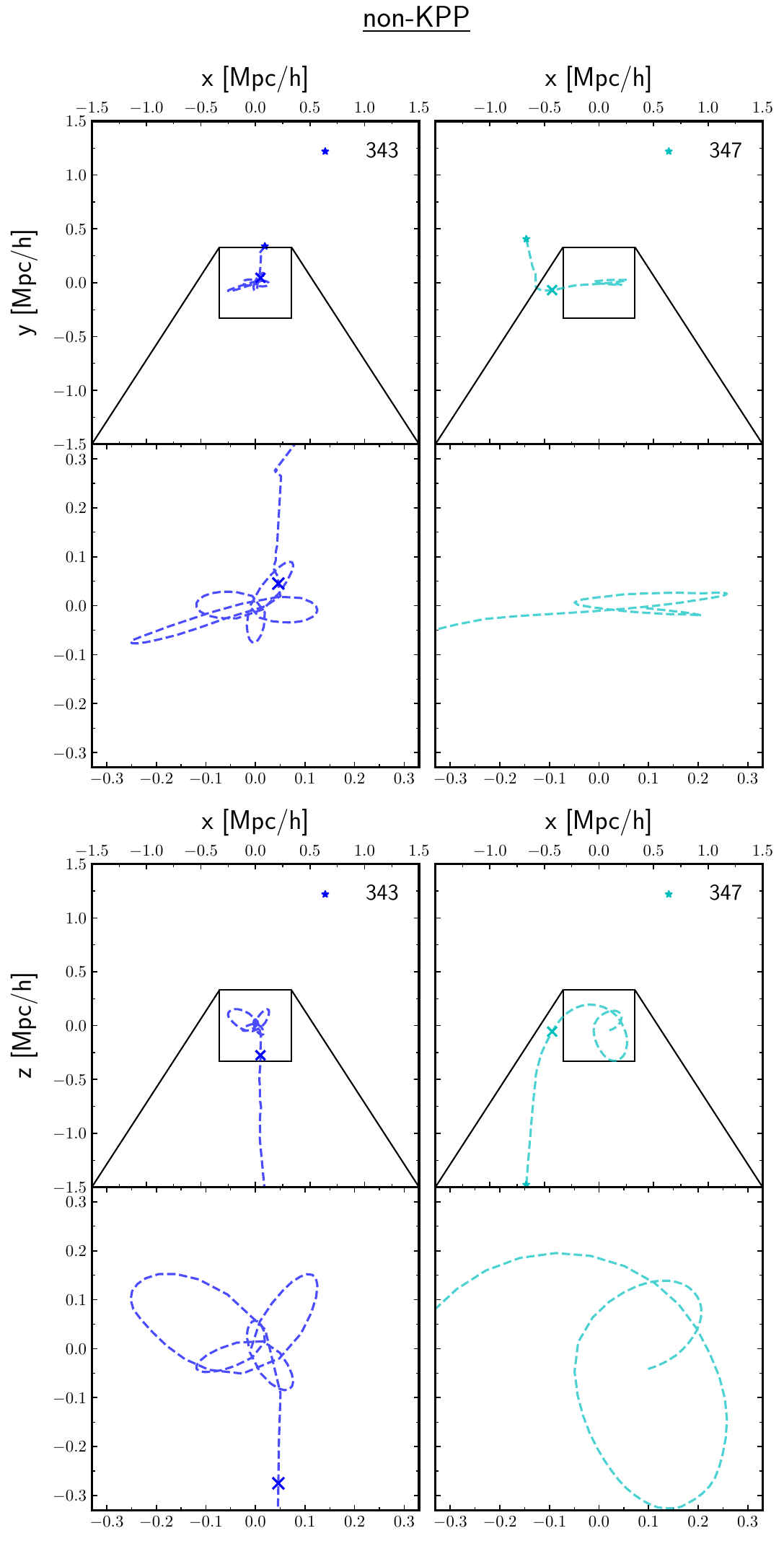}

\caption{Trajectories of four satellites belonging to KPP (leftmost columns) and non-KPP  (rightmost columns) populations  in the Aq-C$^{\alpha}$ system. 
Two  projections are shown for each satellite, either  along the 
$ \hat{e}_3$ principal direction  (two upper  panels) or along the   $ \hat{e}_2$ principal direction at $z=0$  (two bottom  panels). A zoom of the respective central regions is also shown.
Trajectories go from z$_{\rm high}$ until $z=0$, in comoving (box) coordinates with the host galaxy center as origin of the reference system.
The dots in the non-zoomed  panels indicate the position of the proto-satellite (see ID in legend) at $z_{\rm high}$ given in Table \ref{tab:TabData1}. Small crosses on the trajectories mark satellite positions at  T$_{\rm dist,plane}$ (i.e. the collapse timescale for KPP satellites onto the $\hat{e}_3$-structure).
}
\label{fig:Fragm-SLICES_TRAJEC}
\end{figure*}

A  second interesting result Figure \ref{fig:dist-median-perc-aqc} reports on, is that KPP and non-KPP satellites sample differently the two phases mentioned above.
Differences before and along the $\hat{e}_3$-structure collapse are of particular interest here.   
Non-KPP satellites  come from further away than KPP satellites before T$_{\rm dist, plane}$
(as Figure \ref{fig:Fragm-SLICES_TRAJEC}  illustrates), and, on average,  
non-KPP satellite members reach the $\hat{e}_3$-plane later on than KPP satellite members as well. The specific values of T$_{\rm dist, plane}$ for the two satellite populations are given in  Table \ref{tab:TimescaleTable}. 

Additionally, by following (proto-)satellite spatial trajectories, 
we have found that many non-KPP satellites trace mass flows that are mostly perpendicular to the $\hat{e}_3$-plane, converging onto the plane and feeding it, while KPP satellites' 
trajectories tend to  run more obliquely relative to the $\hat{e}_3$-plane. Figure \ref{fig:Fragm-SLICES_TRAJEC} shows some illustrative examples of these two different satellite trajectory behaviors.

As seen in Figure \ref{fig:jorb_ei_AqC_PDEVA}, it takes some time to have orbital poles of KPP satellites aligned with the  $\hat{e}_3$ directions (indeed, T$_{\rm align, e_3} \simeq$ 4.5 Gyr, see subsection \ref{sec:alignOP}). To reach the aligned configuration that roughly keeps stable in time, a necessary condition is that the $\hat{e}_3$ principal direction is frozen. This happens very early (at T$_{\rm dir, e_3} \sim $ 2 Gyr, see Table \ref{tab:TimescaleTable}). 
A second condition is that those not-yet-in-plane incoming satellites are placed in-plane.

Within this scheme, KPP satellite members  would be those placed in-plane from very high redshift, plus those having their orbital poles sucessfully bent while they are incorporated into the $\hat{e}_3$-structure.
This ‘pole-bending’ effect comes about as a result of the particular dynamical evolution of the CW, where satellites are just tracers of its flowing mass elements. Indeed, the trajectories of the KPP satellites  shown in Figure \ref{fig:Fragm-SLICES_TRAJEC} are evidently bent towards the $\hat{e}_3$-wall (approximately the XY plane) when collapsing onto it at T$_{\rm dist,plane}$ (see ``x'' markers).

Non-KPP satellites can  have different origins: i),  those coming from further away than KPP satellites, infalling  onto the $\hat{e}_3$-plane at late times and thus  not having had enough time to complete a full orbit around the host yet, ii),  those with low impact parameters and hence small apocenters, like satellite \# 343  in Figure \ref{fig:Fragm-SLICES_TRAJEC}, and, iii) those satellites with almost perpendicular infall onto the $\hat{e}_3$-plane, but with larger impact parameter, such that their orbital poles are parallel to the plane and aligned with the $\hat{e}_2$ axis (3 out of 21 non-KPP satellites, not classified as a second plane due to their low number), see satellite \# 347 in Figure \ref{fig:Fragm-SLICES_TRAJEC}.



%
Some considerations are in order concerning the effects that satellite motions within the $\hat{e}_3$-structure may have on satellite pole alignments.
As already mentioned, between T$_{\rm dist, plane}$ and T$_{\rm vir}$ the alignments between KPP satellite poles and the $\hat{e}_3$ direction improve somewhat, but to a  much lesser extent  than during the previous period of $\hat{e}_3$-structure formation. Thus, it would seem that, after T$_{\rm dist, plane}$,  the secondary collapse phase leading to the formation and evolution of this prolate structure does not have relevant  effects  on the alignments.
Indeed, as already mentioned, these alignments keep overall constant until z $\sim$ 0 (see Figure \ref{fig:jorb_ei_AqC_PDEVA}), in contrast to early CW walls and filaments, that mostly vanish in favor of halos (see Figure \ref{fig:lagvol}).


\subsection{PDEVA-5004: two KPPs within a prolate CW structure}

In the case of PDEVA-5004, the right panel of Figure \ref{fig:dist-median-perc-aqc} 
indicates that the median distances to the $\hat{e}_3$-plane for the different satellite populations show similar trends
as those mentioned above for the Aq-C$^{\alpha}$ simulation, i.e., two clearly distinguished phases, the first of them of rapid distance decrease. 
The main dissimilarity with the Aq-C$^{\alpha}$ case is that in PDEVA-5004 the $\hat{e}_3$-structure is a prolate structure and not a wall.

Another particularity of satellites in PDEVA-5004 is that, in many cases, proto-satellite trajectories are initially approximately parallel to  the $\hat{e}_3$ principal direction,    and have their respective orbital poles  aligned since very high redshift with the $\hat{e}_1$ principal direction. This is  the origin of the KPP1 satellite system. 

KPP1 and KPP2 satellite groups show different T$_{\rm dist,plane}$ values (see Table \ref{tab:TimescaleTable}): KPP1 satellite members reach, on average, the $\hat{e}_3$-plane
$\sim$ 2 Gyr later than KPP2 satellites, during a phase of the evolution when dynamical events happen rapidly.
%
Again, KPP2 satellite members are those whose trajectories have been succesfully bent by  T$_{\rm dist, plane} \sim $ 3 Gyr (and by the same reasons advocated previously, i.e. the CW dynamics), or those that are already aligned at high redshift.
In addition, KPP1 satellite members come from further away than KPP2 members,
with their poles already clustered in the $\hat{e}_1$ direction, and this clustering is maintained across $\hat{e}_3$-structure collapse.

As for non-KPP satellites, 
similarly to the Aq-C$^{\alpha}$ case, most of them reach the $\hat{e}_3$-plane
with a low impact parameter relative to the host center, or have been recently captured by the host.



\subsection{Timescales}
\label{sec:Timescales}

Different timescales for the galaxy-satellites system formation and evolution have emerged along this work, summarized in Table \ref{tab:TimescaleTable}.
Apart from the halo turn-around, T$_{\rm ta, halo}$, and virialization, T$_{\rm vir}$, timescales (Table \ref{tab:TabData1}) coming from the Spherical Collapse Model, 
we have pointed to and made specific definitions for timescales relative to the LV evolution, such as the Universe age when:
\begin{enumerate}[label = {\roman*)}, wide, labelwidth=!, labelindent=0pt]
    \item the $\hat{e}_3$ principal vector,  corresponding to the direction of the  dominant compression  flow  of matter, gets  its direction fixed, T$_{\rm dir, e_3}$ (see Figure \ref{fig:direc-evol-DM});
    \item the rapid high redshift decrement of the minor principal axis, $c(t)$, stops:  T$_{\rm shape, e_3}$ (see Figure \ref{fig:LVs-Properties_AqC-PDEVA}).    
\end{enumerate}
 We have also focused on timescales involving the (proto-)satellites of the different populations, in which case, when possible, we show the population median values and the 25th - 75th percentiles  of the Universe age when:
 \begin{enumerate}[label = {\roman*)}, resume*]
    \item the i-th satellite becomes aligned  with the  axis $\vec{J}_{\rm stack}$  of maximum  satellite co-orbitation determining  the KPP plane the satellite belongs to, T$_{\rm cluster, Jstack}$ (see Figure \ref{fig:jorb_3Jorb});
    \item the i-th satellite becomes aligned with the  $\hat{e}_3$ principal vector (for satellites in KPP or KPP2) or with the $\hat{e}_1$ eigenvector (satellites in KPP1); see Figure \ref{fig:jorb_ei_AqC_PDEVA} and T$_{\rm align, e_i}$ entry in Table \ref{tab:TimescaleTable};
    \item there is a broad minima in the median vertical distances to the $\hat{e}_3$-plane of the different satellite populations shown in Figure \ref{fig:dist-median-perc-aqc}, T$_{\rm dist, plane}$. From this age onwards, the median distances stay roughly constant (in physical coordinates), and are much lower than the distances before this age is reached.

\end{enumerate}
For the sake of the discussion in this subsection, we now report on two more timescales involving satellite populations as well, i.e., their medians and percentiles of the Universe age when:
\begin{enumerate}[label = {\roman*)},resume*]
    \item the i-th satellite distance to the host center-of-mass is, for the first time, smaller than the respective virial radii at that age  (infall time), T$_{\rm sat, infall}$; and
    \item the distance from the i-th (proto-)satellite to the (proto-)halo center-of-mass  reach their first maximum (in physical coordinates), T$_{\rm sat, apo1}$. This is essentially a turn-around timescale for the $i$-th satellite, marking the beginning of its decoupling from the expansion of the background Universe.  
\end{enumerate}

As discussed in the previous subsections, in the physical process behind satellite orbital pole alignment with the $\hat{e}_3$ principal direction, a timescale stands out in the two zoom-in simulations studied: T$_{\rm dist, plane}$, the collapse timescale for the $\hat{e}_3$-structure. Its values are $\sim$ 3.5 Gyr  for the KPP system in the Aq-C$^{\alpha}$ simulation,
 and  $\sim$ 3 Gyr for the KPP2 system in  PDEVA-5004.
According to Table \ref{tab:TimescaleTable}, T$_{\rm dist, plane}$ is roughly coeval to the clustering timescales, or the Universe age when KPPs are established: T$_{\rm cluster, Jstack}$ or T$_{\rm align, e_3}$. This is expected, because an aligned satellite orbits within the $\hat{e}_3$-structure, thus its distance to the $\hat{e}_3$-plane must be low.
This Table also indicates that T$_{\rm dist, plane}$ is coeval to T$_{\rm shape, e_3}$, a timescale marking -- within the accuracy of their  determinations -- the Universe age when 
the initially strong mass flows normal to the $\hat{e}_3$-structure, which feed it and eventually cause its collapse, slow down.

On the other hand, the infall of satellites onto the halo happens after their capture by the $\hat{e}_3$-structure (particularly so for the Aq-C$^{\alpha}$ simulation),
while the satellites reach first apocenter at times much earlier than this (see Table \ref{tab:TimescaleTable} T$_{\rm sat, infall}$ and T$_{\rm sat, apo1}$ entries).
Thus, none of these processes seem to be related with the origin of KPPs.

Alignments with the principal directions set in very early for KPP members in both simulations. For PDEVA-5004 KPP2 satellites, this happens  earlier than for satellites in the Aq-C$^{\alpha}$ system.
As for PDEVA-5004 KPP1  satellites, they  come with the main mass flow normal to the $\hat{e}_3$-plane, and therefore their clustering (defined by the $\hat{e}_1$ direction) is already established at very high redshift. 
 
To finish, let us mention the timescale for the establishment of the maximum ordered disky motion (circular and in-plane) of satellites in the two simulations analyzed here,  T$_{\rm max, rot}$ (Section \ref{sec:krot}).
We see that the morphological kinematic $\kappa_{\rm rot}$ parameter reaches its maximun shortly after T$_{\rm dist, plane}$, both for the KPP and the KPP2 satellite samples, in coherence with our previous interpretations in this Section.
To complete the scheme, for the KPP1 satellite members the $\kappa_{\rm rot}$ parameter values are high since very high redshift, in coherence with the very early alignments of their orbital poles with the $ \hat{e}_1(t)$ principal direction.




\section{Summary and Conclusions}
\label{sec:SummConclu}


\subsection{Summary}

The aim of this paper is to make a contribution to understanding the origin or physical processes behind  the formation  of persistent, kinematically-coherent planes of satellites (KPPs), i.e., sets of satellites, with fixed identities, co-orbiting around their host galaxy,  whose orbital poles are conserved and clustered  across  long cosmic time intervals, and whose positions form good quality positional planes. 

\citet{Santos-Santos_2023}, Paper III, identified such KPPs in their analyses of  two cosmological, zoom-in hydrodynamical simulations where a system of some 30-35 satellites  orbit around a MW-mass type galaxy with an  extended thin gaseous and stellar disk. 
The two simulations differ in their initial conditions, the subgrid physics, and the methods to integrate both the gravitational and 
the hydrodynamical equations. In Paper III, simulations are analyzed from halo virialization time T$_{\rm vir}$ to $z =  0$.
In both simulations a relatively high fraction of the satellites have been found to be kinematically organized
(a maximum of a $\sim$ 60\%  and of  a $\sim$ 80\%,  along some time intervals, in the  Aq-C$^{\alpha}$ and PDEVA-5004 systems, respectively).

The specific aim of this paper is to elucidate which physical processes cause the satellite orbital angular momentum ($\vec{J}_{\rm orb}$) direction clustering at early times.
Using the same two zoom-in simulations as in Paper III, but extending the analyzed period back in time until 
$\zhigh$ = 8.45 and 10.00  for the  Aq-C$^{\alpha}$ and PDEVA-5004 systems, respectively,   we focus on the overall  evolution of the Cosmic Web (CW) around the satellite-to-be and galaxy-to-be objects from $\zhigh$  until $z = 0$. 
We therefore follow the progenitors of the low redshift satellites and  host systems well within the fast phase of the system assembly,  and within the local (i.e., around the forming system) CW they are embedded in.

By following satellites back in time, we find that, in most cases,  their progenitors 
gain specific orbital angular momentum magnitude  as predicted by the Tidal Torque Theory from T$_{\rm uni} \sim 2$ Gyr to $\sim$ 4 Gyr. 
The directions of the orbital angular momentum (i.e., the so-called orbital poles) of KPP  satellite progenitors  are also conserved from very early, while this is not the case for many satellites outside KPPs (Figure \ref{fig:sJ-zoom-pole-cons}).  
Our  analysis here indicates  that clustering of orbital poles 
occurs  already at high redshift, see Table \ref{tab:TimescaleTable}. 
An analysis by means of the morphological kinematic $\kappa_{\rm rot}$ parameter \citep[see, for example,][]{Sales:2012}  indicates that, from very early times, 
the collections of KPP satellites represent, kinematically, ``disky'' systems,
with a high fraction of their kinetic energy coming from in-plane and almost circular motion within KPPs, while systems outside these structures are more spheroidal-like (Figure \ref{fig:Krot-median-satellites}). 

	To elucidate how the aforementioned  clustering came about, we analyze satellite pole evolution as part of the Cosmic Web they are embedded in, through the LV  deformation method \citep[see][]{Robles:2015}. 
For each simulation, we mark the particles that at $\zhigh$ are within an sphere of radius  $R_{LV}$ = $K\cdot r_{\rm vir, z=0}/(1 + \zhigh)$ ($K$=15 and 20) centered at the protogalaxy center, and follow their trajectories forward in time up to $z =  0$.        
The volumes these particles span at each time, are the so-called Lagrangian Volumes (LVs).
We analyze the evolution of their principal directions $ \hat{e}_i(t)$, with $i = 1, 2, 3$,
and their principal axes $a(t) >  b(t) > c (t)$ through the reduced Tensor of Inertia (TOI) method \citep{Cramer}.  

The $a(t), b(t), c(t)$ functions inform  us about the shape deformations of the LV, including how quickly they happen. 
The general result is that while $a(t)$ grows, $c(t)$ decreases, in most cases monotously, but with some stagnation periods (Figure \ref{fig:LVs-Properties_AqC-PDEVA}). The $b(t)$ axis keeps roughly constant in PDEVA-5004, and decreases in Aq-C$^{\alpha}$ after T$_{\rm uni}\sim$ 4 Gyr.
 As for axes ratios, $c(t)/a(t)$   decreases rapidly up to T$_{\rm uni}\sim$ 4 Gyr in the Aq-C$^{\alpha}$ LV, 
causing a quick  flattening of its  initially spherical shape into a wall-like CW structure (Figure \ref{fig:lagvol}). Then, the decrement of the $b(t)/a(t)$ ratio  takes over,  transforming the LV shape from oblate to triaxial, and finally prolate. Ratio changes in the PDEVA-5004 LV are such  that its shape is always triaxial.
The principal directions also freeze out, meaning that the direction of overall maximum compression $ \hat{e}_3$
does not change after its freezing out, within a threshold, T$_{\rm dir,e_3}$ and, consequently, from this moment onwards, the overall maximum compression takes place along a fixed direction (Figure \ref{fig:direc-evol-DM}).
All three principal directions freeze out  very early (T$_{\rm freeze} \simeq 4$ and 2 Gyr for Aq-C$^{\alpha}$ and PDEVA-5004, respectively).
In this way, we witness the emergence of a kind of global LV `skeleton', such that for  T$_{\rm uni} >$ T$_{\rm freeze}$,  overall   anisotropic mass rearrangements  of LV particles  occur along  fixed directions.

To elucidate the role that the local CW development around the forming system has at driving pole clustering, 
we have analyzed the alignments between  the LV principal directions and  the satellite orbital poles across cosmic time (Figure \ref{fig:jorb_ei_AqC_PDEVA}).
We find that, for KPP satellites in the Aq-C$^{\alpha}$ system,  a clear alignment signal  stands out with the $ \hat{e}_3$ axis  after T$_{\rm align, e_3} \sim$ 4.5 Gyr (see Table \ref{tab:TimescaleTable}),  i.e., orbital poles of KPP satellites tend to be  parallel to the LV's direction of maximum overall compression. 
Some satellites in the KPP show alignments as early as at T$_{\rm uni} \sim$ 2 Gyr. For those that do not, the orbital pole are bent efficiently between T$_{\rm uni}\sim$ 2 - 4.5 Gyr. This is roughly coeval to the timescale for $\hat{e}_3$-structure collapse.
Satellites outside KPP structures do not show any particular alignments with any LV principal direction.

In the PDEVA-5004 system, the pole alignments signal of KPP2 satellites with the $ \hat{e}_3$ axis  
 after T$_{\rm align,e_3} \sim$ 3.5  Gyr is even clearer than in the Aq-C$^{\alpha}$ case.   
KPP1 satellite poles are well aligned with the $ \hat{e}_1$ axis since at least  T$_{\rm uni}\simeq$ 2.0 Gyr.
No alignment signals show up for satellites not belonging to KPPs.

We study the evolution of proto-satellite mass elements in relation to the evolving local CW (Figures \ref{fig:Proj-Dens-Sat} and \ref{fig:dist-median-perc-aqc}). Our findings show that KPP satellites aligned with $\hat{e}_3$ present closer, more oblique trajectories relative to the $\hat{e}_3$-plane, allowing for ``pole bending'' by the same forces and torques that drive the evolution of the local CW dynamics (Figure \ref{fig:Fragm-SLICES_TRAJEC}). Satellites with poles aligned with the $\hat{e}_1$ axis show trajectories parallel to $\hat{e}_3$ and are not bent. Finally, non-KPP satellites can present different origins, with many showing low impact parameters relative to the host center and hence high probability of suffering disturbing phenomena.

\subsection{Some comparisons with other works}
\label{sec:comparisons}

To our knowledge, this paper represents  the first time that the effects of the CW as a driver of the satellite orbital pole organization into KPPs is analyzed in some detail through numerical simulations and where comparison with previous results, including observational ones, can be  easily made.

A few works have attempted to address the characteristics and evolution of co-orbiting satellite planes, or their origin as connected to the large scale structure they are embedded in. 

For example, \citet{Shao19} used the EAGLE-100 volume to identify “MW-like-orbit” satellite planes, i.e., narrow planes formed by the 11 most massive satellites around MW-mass halos and where 8 of them show a high degree of co-planarity (a small dispersion of their orbital poles) at $z=0$. In that paper, their aim was to look for alignments with the principal directions  of the host halos, resulting from a TOI analysis.
%
%
%
%
%
They noted that the degree of co-orbitation of their subsets of 8 coherent satellites selected at $z=0$ is best at present than at earlier times, and suggest it is driven by halo torques after infall. 
Our findings are consistent with theirs in that the collimation and clustering of satellite orbital poles in two KPPs improves with time (see alignments with the $ \hat{e}_3$ direction in both simulations, Figure \ref{fig:jorb_ei_AqC_PDEVA}). 
Finally, it is worth noting that they found a wide variety of times at which these 8 MW-like-orbit satellites started to show co-orbitation, with some setting in early, 
while  others do later on (see the wide range implied by  percentiles in T$_{\rm align, e_3}$ in Table \ref{tab:TimescaleTable}).
The higher fraction of satellites  that are established much later on relative to our results could come from other possible channels for orbital pole clustering enhancement such as LMC-like group infall or tides from aspherical halos.

In a more general perspective, our results are also consistent with --and provide an explanation to-- those from \citet{Libeskind14,Dupuy22} who found a preferred direction of subhalo infall onto halos. 
These works show that subhalos 
are mainly incorporated onto halos along a direction that is contained within the plane orthogonal to the direction of fastest collapse, and that aligns with the spines of filaments. Following these predictions, \citet{Libeskind15,Libeskind19} tested the possible alignments between the observed satellite planes  in the Local Universe and the principal directions of the cosmic density field as reconstructed from the CosmicFlows-2 peculiar velocity survey \citep{Tully13}.
%
Similarly, \citet{Xu23} also suggest a possible connection between the presence of a rotating plane of satellites in TNG50 and the large-scale sheet structure it is embedded in.

Our work shows that, in the two simulations analyzed in this Paper, kinematically-coherent satellites in fact gain their common dynamics \textit{much before} 
they reach the halo, by tracing the mass flows as mass collapses into the CW elements \citep{Zeldovich70,Shandarin89}.  
We emphasize, therefore, that here it is but  the initial step of anisotropic mass collapse, 
together with the initial location of proto-satellites relative to the skeleton of CW collapsed structures, which leaves an imprint on the dynamics of satellites.
%

Indeed, we find that --differently to suggestions from previous works \citep{Goerdt13,Buck15,Ahmed17}--, kinematic planes are not only driven by filamentary infall, as KPP satellites do not always reach their stationary positions  via one or a few strong filaments.

We note that the principal directions of collapse identified in the previous works 
result from velocity shear tensor analyses at $z$ = 0.
Indeed, most methods to analyze the CW evolution and classify its elements \citep[see, e.g.][and references therein]{Hoffman2012,Cautun:2013,Libeskind18} are local ones, where the tensorial tools used are defined on points, needing a smoothing procedure to enable calculation.

Conversely, to study the evolution of the local environment of  sites   where  galaxy systems are to form,
our perspective is rather global.
We use here  a simple  method which tracks  the average large-scale deformations of an initially spherical Lagrangian Volume.
The method provides the accumulated deformations the LV suffers   along time intervals  between the different snapshots the simulation provides.
This method is simpler than the usual local, tensorial methods  in that
the LV evolution is described through 3 principal directions (that happen to freeze out at very early times) and 3 time functions,
the principal axes $a(t)$, $b(t)$, $c(t)$, from which only two are independent. 
The method accurately catches the evolution of the local environment of galaxy
formation sites.
In our analysis here, the method  singles out an overall  direction of maximum compression of the mass flows whose relevance here has already been mentioned.
In practice, this direction of maximum compression is the same as those  returned by the usual methods, as the velocity shear tensor analysis\footnote{Please  note the different nomenclature used within both frameworks, where the direction of fastest collapse corresponds to $\vec{e}_1$ from the velocity shear tensor and to $\vec{e}_3$ from the inertia tensor (minor axis of the ellipse).}.
While the velocity shear tensor is a more appropriate scheme when trying to detect and classify cosmic web structures \citep[see][]{Cautun:2013,Libeskind18}, both formalisms allow to identify the main directions of mass flows.


\subsection{Conclusions}

These are the conclusions of this work concerning the origin of KPPs, according to the two simulations analyzed in this Paper: 
\begin{enumerate}[wide, labelwidth=!, labelindent=0pt]
    \item The formation of KPPs is closely related to the early anisotropic collapse of mass in $\Lambda$CDM that drives the evolution of the CW: the same physical processes behind the emergence of the CW at high redshift are behind the emergence of clustering of (proto-)satellite orbital poles, that is, behind the formation of KPPs at high redshift. The initial location of proto-satellites relative to the sites of early CW collapse also plays a role.
    \item A timescale stands out for the establishment of KPP satellite orbital pole clustering: T$_{\rm dist, plane}$, the Universe age when satellites' distances to the plane defined by the direction of overall compression of the local mass distribution (i.e. $\hat{e}_3$ direction) become overall roughly constant. This is similar to  a collapse event. This event occurs well before the median timescale for satellite infall  onto their host halo.
%
%
    \item KPP member satellites characterized by pole alignments with the $\hat{e}_3$ principal direction are those already aligned at high redshift, plus those whose orbital poles have been succesfully bent as they are incorporated into the $\hat{e}_3$-structure when it collapses. The latter's trajectories at high redshift tend to be oblique relative to the $\hat{e}_3$-plane. This is the case of the so-called Aq-C$^{\alpha}$ and PDEVA-5004 KPP2 groups. Additionally, in the case of Aq-C$^{\alpha}$ KPP satellites, they collapse earlier onto the $\hat{e}_3$-structure, and are closer to the $\hat{e}_3$-plane at given times, than non-KPP satellites.
    \item KPP satellite orbital pole alignments occur not only along the direction of maximum compression $\hat{e}_3$, but can also appear along the other principal directions. Such is the case of a sizeable subset of satellites in the PDEVA-5004 simulation (the KPP1 system), whose orbital poles are aligned with the $\hat{e}_1$ direction, and of some satellites in the Aq-C$^{\alpha}$ simulation, with orbital poles aligned with the $\hat{e}_2$ eigenvector.
    \item In PDEVA-5004 the CW dynamics leads to a triaxial mass distribution (rather than to a wall-like structure as in Aq-C$^{\alpha}$). Those PDEVA-5004 satellites characterized by pole alignments with the $\hat{e}_1$ tend to have trajectories that, at high redshift, follow the direction of overall maximum compression. These satellites do not suffer from orbital pole bending when the $\hat{e}_3$-structure collapses.

\end{enumerate}

We would like to emphasize that just two simulations have been analyzed. However, these two simulations differ in multiple aspects. In Section \ref{sims} we presented the differences between the two simulations regarding their hydrodynamics,
and we pointed out that the initial conditions and the subresolution physics models  differ as well.
The possibility of KPP formation in both cases implies that their origin must be driven by the more fundamental common physical processes of structure evolution in a $\Lambda$CDM cosmological context, and less dependent on the details of galaxy formation modeling. 
We have shown not only that KPPs can form in $\Lambda$CDM simulated disk galaxy systems, but, importantly, that their existence is a natural consequence of $\Lambda$CDM's prediction for large-scale mass flows at high redshift shaping the local Cosmic Web structure. In the two zoom-in simulations analyzed here, KPPs are the result of the same dynamics acting on proto-satellites' mass elements  placed at particular locations and/or endowed with particular kinematic characteristics.

We note that other channels for KPP enhancement at low redshift are possible as well, for example through the late capture of a satellite with its own system of subsatellites (see Paper III). On the other hand, satellite interactions within the inner regions of halos could destroy kinematic coherence.

Finally, we want to stress that our scientific conclusions are an interpretation of results from only two simulations, which exhibit a correlation between LVs and KPPs.
Further work involving extending our analysis to a broader sample of simulations is therefore needed in order to assess the frequency of finding KPPs, and to robustly conclude that this is a generic feature of KPPs in $\Lambda$CDM. In particular, analyses of large-volume simulations of galaxy formation might shed light on how frequently the different channels for satellite plane formation appear throughout cosmic evolution.


\section*{Acknowledgements}

We thank the Ministerio de Ciencia e Innovación (Spain)  for financial support under Project grant PID2021-122603NB-C21.
M.G.M. thanks MINECO/FEDER funding (Spain) through a FPI fellowship associated to PGC2018-094975-C21 grant.
I.S.S. acknowledges support by the European Research
Council (ERC) through Advanced Investigator grant to C.S. Frenk, DMIDAS (GA 786910).
S.E.P. acknowledges support from MinCyT (Argentina) through BID PICT 202000582.
P.B.T. acknowledges partial
funding by Fondecyt 1200703/2020 (ANID) and CATA-Basal-FB210003 project.%
M.A.G.F acknowledges financial support from the
Spanish Ministry of Science and Innovation through the project
PID2020-114581GB-C22.
This work used the  Geryon cluster (Pontificia Universidad de Chile). We used a version of Aq-C-5 that  is part of the CIELO Project run in Marenostrum (Barcelona Supercomputer Center, Spain), the NLHPC (funded by ECM-
02) and Ladgerda cluster (Fondecyt 12000703).
This project has received funding from the European Union   Horizon 2020 Research and Innovation Programme under the Marie Skłodowska-Curie grant agreement No 734374- LACEGAL.

\bibliography{archive_LV_cruzado_MGM_segu}{}
\bibliographystyle{aasjournal}


\appendix
\restartappendixnumbering

\section{Evolution  of   the  principal directions}
\label{app:K_comparisons}

\begin{figure}
\centering
\includegraphics[width=0.9\linewidth]{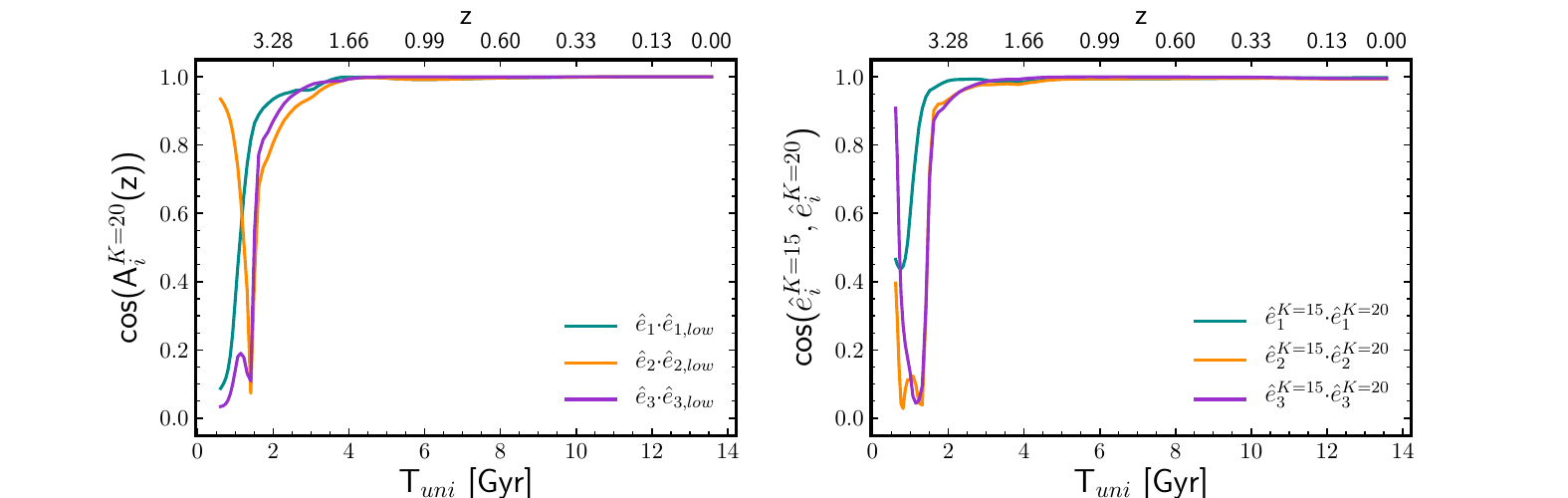}
%
%
\caption{Top panel: Evolution of the cosine of the angle $A_i$ formed by the eigenvectors $e_{i}(z)$ and $e_{i}(z=0)$ with $i$ = 1,2,3 for the  Aq-C$^{\alpha}$ simulation for a LV $K=20$ scale, that is, $R_{\rm LV}$ = 20$\cdot r_{\rm vir, z=0}/(1 + \zhigh)$.
Bottom panel: relative orientations of the principal directions of  LVs tracing  the $K=15$ and $K=20$ scales.
Upper horizontal axes give the redshift scale, while the lower ones stand for the Universe age T$_{\rm uni}$.}
\label{appendix:direc-evol-DM}
\end{figure}

In this Appendix we test how the principal directions of compression behave when we change the LV scale, $R_{LV} = K\times r_{\rm vir, z=0}/(1+z_{\rm high})$.
In the top panel of Figure \ref{appendix:direc-evol-DM}, we see that the timescale for principal direction fixing when using a scale of $K=20$ (involving a volume increase of a  $\sim 2.4$ factor compared to the fiducial $K=15$) behaves identically as for $K=15$, as none of the principal directions change after  T$_{\rm uni} \sim 4$ Gyr $\equiv$ T$_{\rm freeze}$ either for $K=15$ or $K=20$. The same is true in the case of the PDEVA-5004 after T$_{\rm uni} \sim$ 2 Gyr $\equiv$ T$_{\rm freeze}$, except for the very small changes  around T$_{\rm uni} \sim$ 6 Gyr.

To deepen into  these results, we have also analyzed how the principal directions of  LVs at different scales are oriented with respect to each other.  For the Aq-C$^{\alpha}$ simulation these principal directions after T$_{\rm uni} \sim$ 4 Gyr  are essentially the same for either scale, see Figure \ref{appendix:direc-evol-DM}, bottom panel. 
For PDEVA-5004 this is also true from the very  beginning of our analysis, more accurately for the $\hat{e}_{3}$ directions.
Finally, it is worth to note that, however,  when using a $K=10$ value, only one axis freezes out that early, while for the other two it takes a longer time, as they become frozen within a 10\% by  T$_{\rm uni} \sim $ 6 Gyr. This is due to the fact that a shorter scale tracing of the density field around the galaxy-to-be formation site gives a LV dominated by denser mass elements, where more abrupt/ complex dynamic processes take place.
Therefore,  $K=10$ LVs at high $z$ are dominated, in both simulations, by early activity in their central regions and are thus not suited for our purposes here.



\end{document}